\documentclass[twocolumn]{aastex631}

\newcommand{\arii}{\hbox{[Ar$\,${\scriptsize II}]}}
\newcommand{\ariii}{\hbox{[Ar$\,${\scriptsize III}]}}

\newcommand{\oiii}{\hbox{[O$\,${\scriptsize III}]}}

\newcommand{\niII}{\hbox{[Ni$\,${\scriptsize II}]}}

\newcommand{\naiii}{\hbox{Na$\,${\scriptsize III}}}

\newcommand{\neii}{\hbox{[Ne$\,${\scriptsize II}]}}
\newcommand{\neiii}{\hbox{[Ne$\,${\scriptsize III}]}}
\newcommand{\neiv}{\hbox{[Ne$\,${\scriptsize IV}]}}
\newcommand{\nev}{\hbox{[Ne$\,${\scriptsize V}]}}
\newcommand{\nevi}{\hbox{[Ne$\,${\scriptsize VI}]}}

\newcommand{\feii}{\hbox{[Fe$\,${\scriptsize II}]}}
\newcommand{\feviii}{\hbox{[Fe$\,${\scriptsize VIII}]}}

\newcommand{\siii}{\hbox{[S$\,${\scriptsize III}]}}
\newcommand{\siv}{\hbox{[S$\,${\scriptsize IV}]}}

\newcommand{\ha}{\hbox{H$\alpha$}}
\newcommand{\hb}{\hbox{H$\beta$}}
\newcommand{\paa}{\hbox{Pa$\alpha$}}

\newcommand{\mgv}{\hbox{[Mg$\,${\scriptsize V}]}}

\newcommand{\kms}{km\,s$^{-1}$} 
 
\newcommand{\um}{$\mu$m}
\newcommand{\eden}{cm$^{-3}$}

\newcommand{\ergs}{erg\,s$^{-1}$}
\newcommand{\ergscm}{erg\,s$^{-1}$\,cm$^{-2}$}

\newcommand\jwst{\emph{JWST}}
\newcommand\spitzer{\emph{Spitzer}}

\newcommand\ifsfit{\texttt{IFSFIT}}

\newcommand\qtdfit{\texttt{q3dfit}}
\newcommand\questfit{\texttt{questfit}}

\shorttitle{\jwst\ PDS 456}
\shortauthors{Seebeck et al.}

\begin{document}

\title{Combined \jwst-MUSE Integral Field Spectroscopy of the Most Luminous Quasar in the Local Universe, PDS 456}

\author[0000-0002-4014-9067]{Jerome Seebeck}
\affiliation{Department of Astronomy, University of Maryland, College Park, MD 20742, USA}

\author[0000-0002-3158-6820]{Sylvain Veilleux}
\affiliation{Department of Astronomy, University of Maryland, College Park, MD 20742, USA}
\affiliation{Joint Space-Science Institute, University of Maryland, College Park, MD 20742, USA}

\author[0000-0003-3762-7344]{Weizhe Liu}
\affiliation{Department of Astronomy, Steward Observatory, University of Arizona, Tucson, AZ 85719, USA}

\author[0000-0002-1608-7564]{David S.N. Rupke}
\affiliation{Department of Physics, Rhodes College, 2000 N. Parkway, Memphis, TN 38112, USA}
\affiliation{Zentrum für Astronomie der Universität Heidelberg, Astronomisches Rechen-Institut, Mönchhofstr 12-14, D-69120 Heidelberg, Germany}

\author[0000-0002-0710-3729]{Andrey Vayner}
\affiliation{Department of Physics and Astronomy, Bloomberg Center, Johns Hopkins University, Baltimore, MD 21218, USA}

\author[0000-0003-2212-6045]{Dominika Wylezalek}
\affiliation{Zentrum für Astronomie der Universität Heidelberg, Astronomisches Rechen-Institut, Mönchhofstr 12-14, D-69120 Heidelberg, Germany}

\author[0000-0001-6100-6869]{Nadia L. Zakamska}
\affiliation{Department of Physics and Astronomy, Bloomberg Center, Johns Hopkins University, Baltimore, MD 21218, USA}

\author[0000-0002-6948-1485]{Caroline Bertemes}
\affiliation{Zentrum für Astronomie der Universität Heidelberg, Astronomisches Rechen-Institut, Mönchhofstr 12-14, D-69120 Heidelberg, Germany}



\begin{abstract}
Fast accreting, extremely luminous quasars contribute heavily to the feedback process within galaxies. While these systems are most common at cosmic noon ($z\sim2$), here we choose to study PDS 456, an extremely luminous ($L_{bol}\sim 10^{47}$ \ergs) but nearby ($z\sim0.185$) quasar where the physics of feedback can be studied in greater detail. We present the results from our analysis of the \jwst\ MIRI/MRS integral field spectroscopic (IFS) data of this object. The extreme brightness of PDS 456 makes it challenging to study the extended emission even in this nearby object. MIRI/MRS instrumental effects are mitigated by using complementary NIRSpec and MUSE IFS data cubes. We show clear evidence of a multiphase gas outflow extending up to 15 kpc from the central source. This includes emission from warm molecular (H$_2$ $\nu$ = 0 $-$ 0 and 1 $-$ 0) and ionized (e.g. \paa, \oiii, \neiii, \nevi) gas with typical blueshifted velocities down to $-500$ \kms. We are also able to probe the nuclear dust emission in this source through silicate and PAH emission features but are unable to spatially resolve it. Our results are consistent with this powerful quasar driving a radiatively driven wind over a broad range of distances and altering the ionization structure of the host galaxy.
\end{abstract}



\section{Introduction} 
\label{sec:intro}

Most massive galaxies host a supermassive black hole at their centers \citep[e.g.][]{Kor2013, Abu2024}. This black hole, although a small fraction of the galaxy mass, can have a significant impact on the rest of the galaxy through jets, winds, and radiation during active phases \citep[e.g.][]{Kin2003, Vei2005, Fab2012, Wag2012, Kin2015, Fio2017}. When active, the black hole is referred to as an active galactic nucleus (AGN), and the process by which it impacts its host galaxy is known as feedback. The most luminous AGN ($L_{bol} \gtrsim 10^{45}$ \ergs) are known as quasars and their high luminosities allow us to see them at farther distances than other AGN. 

Cosmological simulations have recently shown that feedback from these central supermassive black holes is vital in properly explaining numerous galaxy properties \citep[e.g.][]{Vog2014, Ric2016, Vei2020}. This includes, but is not limited to, galaxy morphology \citep[e.g.][]{Dub2016, Cho2018}, the interstellar medium \citep[e.g.][]{Hop2016, Dav2019}, the circumgalactic medium \citep[e.g.][]{Tum2017}, supermassive black hole growth \citep[e.g.][]{Vol2016, Hop2016}, and star formation quenching \citep[e.g.][]{Zub2012, Pon2017}. The fastest, most powerful winds are generally found in the most luminous quasars \citep[e.g.][and references therein]{Fab2012, Vei2020}. These luminous, fast-wind quasars are most common at $z\sim 2$ \citep[e.g.][]{Zak2016} at the peak of supermassive black hole accretion and star formation rate (SFR). Thus, we predict significant AGN feedback at $z\sim 2$, but direct observational evidence of this feedback has been limited \citep[e.g.][]{Vei2020}.

There are two primary modes of AGN feedback. The first one is known as the ``radiative" or ``quasar" mode, and is expected to dominate in highly luminous, fast-accreting AGN, $L_{\mathrm{AGN}}/L_{\mathrm{Edd}}\geq 10^{-3}$, where $L_{\mathrm{Edd}}$ is the Eddington luminosity. In this mode, the AGN imparts radiative feedback, through emitted photons which heat and ionize the surrounding gas, and mechanical feedback, where radiation pressure imparts momentum on the surrounding gas creating galactic winds. The second mode is the ``kinetic" or ``radio" mode which is expected to dominate at lower Eddington ratios where the radiation pressure is not as significant, $L_{\mathrm{AGN}}/L_{\mathrm{Edd}}\leq 10^{-3}$. In this mode, relativistic jets push gas out of the innermost regions of the galaxy, or at least disturb it, to prevent efficient star formation. Both of these mechanisms have been supported by recent observational evidence (see, e.g., \citealt{Vei2020}; \citealt{Har2024}, and references therein). 

PDS 456 is an extremely powerful local quasar discovered fairly recently \citep{Tor1997}. It is the most luminous quasar in the nearby ($z \la 0.2$) universe with $L_{bol}\sim 10^{47}$ \ergs.  With an estimated black hole mass of $\sim$ (1-2)$\times$ $10^9 M_{\odot}$ \citep{Nar2015}, the implied accretion rate relative to Eddington is $L_{AGN}/L_{Edd} \geq 0.4$ \citep{Nar2015}. More recent observations estimate a mass of $1.70_{-1.15}^{+0.04}$ $\times$ $10^{8} M_{\odot}$, with a implied accretion rate relative to Eddington of 4.64 \citep{GRA2024}. This makes PDS 456 a compelling analog to powerful AGN at the peak of quasar activity, $z\sim 2$. PDS 456 has been extensively studied and shows evidence of outflows at all scales and wavelengths. On the closest scales ($\sim$ 0.001 pc), there is clear evidence of a powerful ``ultra-fast outflow", detected in the hard X-rays ($v_{\mathrm{HX}}\sim 0.25-0.34c$; \citealt{Nar2015}), soft X-rays ($v_{\mathrm{SX}}\sim 0.17-0.27c$; \citealt{Gof2014, Ree2016}), and far ultraviolet ($v_{\mathrm{UV}}\sim 0.30c$; \citealt{Ham2018}). On $\sim 1$ pc scales, UV emission lines have shown evidence for broad winds, decelerated ($v\sim-5000$ \kms) from the nuclear ultra-fast outflow \citep{Obr2005}. Modeling of the spatially resolved \paa\ emission has also shown evidence for an outflow-dominated broad line region at $\sim$ 0.25 pc \citep{GRA2024}. Further out, in the 1-10 kpc regime, ALMA and MUSE data show evidence for a slower ($v\sim-250$ \kms) co-spatial outflow of molecular and ionized gas \citep{Bis2019, Tra2024}. The ubiquity of outflows on all scales in PDS 456 makes it an excellent experimental case study for AGN feedback. 

Over the years, integral-field unit spectroscopy (IFS) has been an extremely useful tool in mapping galaxy-scale extended outflows from local quasars and other galaxies \citep{Nes2008, Rup2011, Liu2013a, Liu2013b, Sin2013, Car2015, Cre2015, Fin2017, Rup2019}. However, the main challenge in luminous quasars, even nearby ones like PDS 456, lies in separating the bright quasar light from the extended host galaxy emission. The excellent spatial resolution and stable point spread function (PSF) of JWST at near- and mid-infared (NIR and MIR) wavelengths have been a true game-changer for this research \citep[e.g.][]{Wyl2022, Vay2023, Vei2023, Rup2023a}. 

In this paper, we analyze the optical, near-infrared, and mid-infrared emission of PDS 456 to constrain AGN feedback in this luminous quasar. In Section \ref{sec:obs_redux}, we discuss the observations and reduction of the \jwst\ and MUSE data. We discuss our analysis methods, including the use of the IFS analysis software package \qtdfit, in Section \ref{sec:analysis}. In Section \ref{sec:results} we present the results of our analysis of both extracted spectra and IFS kinematic maps. We discuss the impact of PDS 456 on the host galaxy and the extent of the multiphase outflow in Section \ref{sec:discussion}. We summarize our conclusions in Section \ref{sec:conclusion}

For this paper, we adopt the flat $\Lambda$CDM cosmology: $H_0$ = 70 \kms\ Mpc$^{-1}$, $\Omega_m = 0.3$ and $\Omega_{\Lambda} = 0.7$. For PDS 456, we use a redshift of $z$ = 0.1850 $\pm$ 0.0001, derived by \citet{Bis2019} from the ALMA CO (3-2) data cube. This gives a luminosity distance $D_L=$ 0.8983 Gpc and a physical scale of 1\arcsec\ = 3.101 kpc. All emission lines are identified by their wavelengths in air (e.g., \oiii\ $\lambda$5007 \AA), but all wavelength measurements are performed on the vacuum wavelength scale. 

\section{Observations and Data Reduction} 
\label{sec:obs_redux}

\subsection{\jwst\ MIRI}
\label{subsec:miri}
\begin{figure}
    \centering
    \includegraphics[width=.47\textwidth]{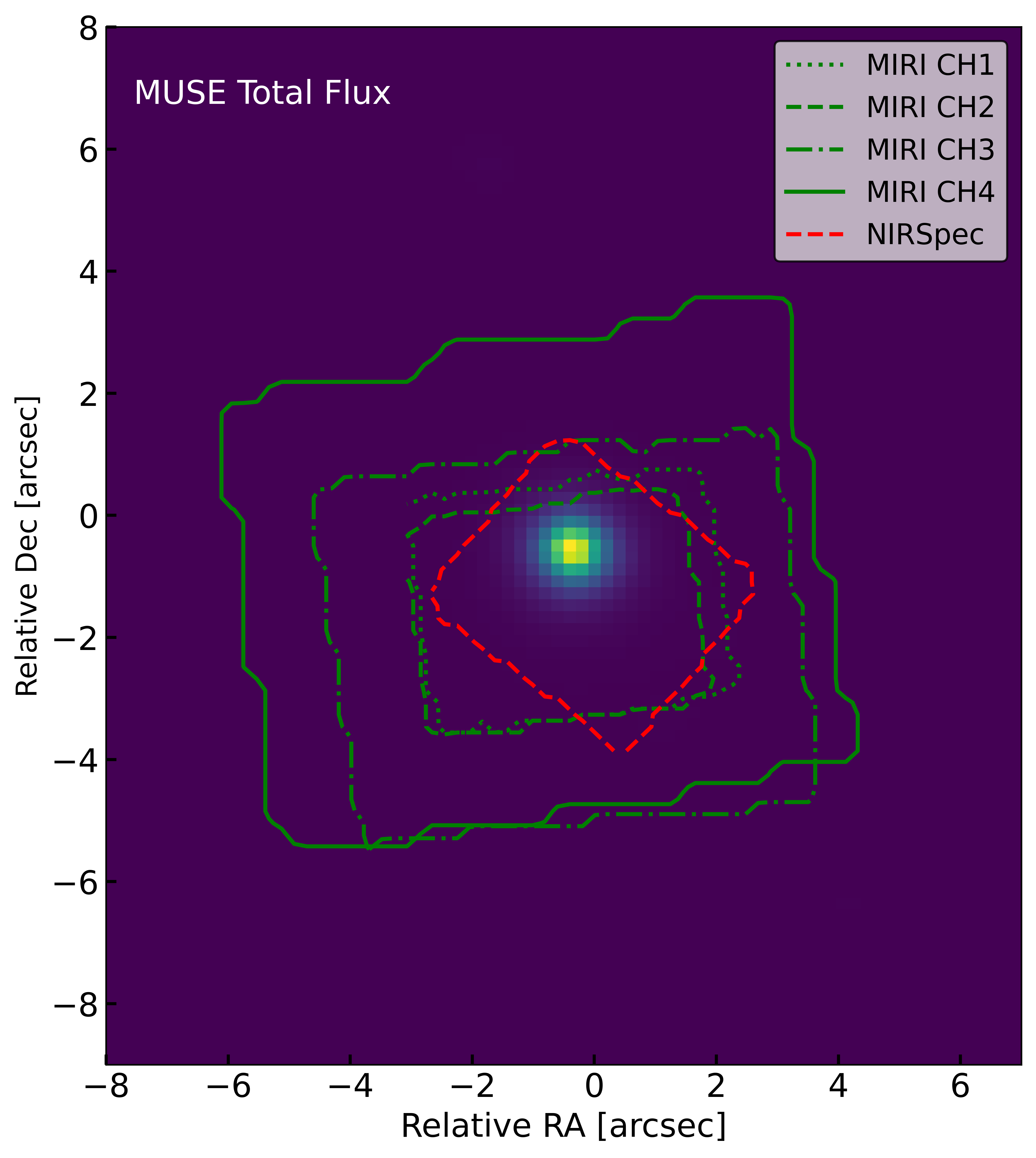}
    \caption{Footprints of the MIRI channels (green) and NIRSpec data (red) plotted over that of the MUSE data, which has been cut down to 14\arcsec\ $\times$ 16\arcsec.}
    \label{fig:footprints}
\end{figure}

PDS 456 was observed with \jwst\ on April 2, 2023 using the Medium-Resolution Spectrometer (MRS) mode of the Mid-InfraRed Instrument \citep[MIRI;][PID 2547, PI Veilleux]{Wri2023, Arg2023}. All of the grating settings, Short, Medium, and Long, were used to achieve the full wavelength range of MIRI (4.9 to 27.9 $\mu m$). The spectral resolution of MIRI ranges from 8 \AA\ in channel 1 to 60 \AA\ in channel 4, corresponding to 30-85 \kms\ across the entire MIRI wavelength range. A 4-point extended dither pattern was used to help remove background contamination and reduce undersampling. The detector footprints of each MIRI channel are shown in Figure \ref{fig:footprints}. 

We reduced these data with \jwst\ v1.13.4 \citep{Bus2022, Bus2024} and v11.17.16 of the Calibration Reference Data System (CRDS). This version of the pipeline shows significant improvements in noise reduction and outlier detection in comparison to older reductions. There are some small issues with large spectral dips in certain regions of the continuum but these are fairly well contained and do not significantly overlap with any spectral lines. The reduction is done with the MIRI pipeline sample notebook publicly available on the Space Telescope github (\href{https://github.com/STScI-MIRI/MRS-ExampleNB/blob/main/Flight_Notebook1/MRS_FlightNB1.ipynb}{MRS\_FlightNB1.ipynb}) with 2D background subtraction and 2D residual fringing steps turned on and the rest of the notebook in its default state. These steps noticeably improve the overall noise and fringing in fully reduced cubes. 

We have additionally attempted other modifications of the pipeline such as: modifying default outlier detection and additional residual fringing. However, these modifications show no major impact on the data and sometimes reduce its quality, thus we did not include them. The \jwst\ reduction pipeline is quickly evolving and has improved greatly over the past year. Many individuals have made modifications to their versions of the pipeline to reduce their data, however, with the speed at which it is improving, we have opted to keep it close to the default settings. 

\begin{figure*}
    \centering
    \includegraphics[width=\textwidth]{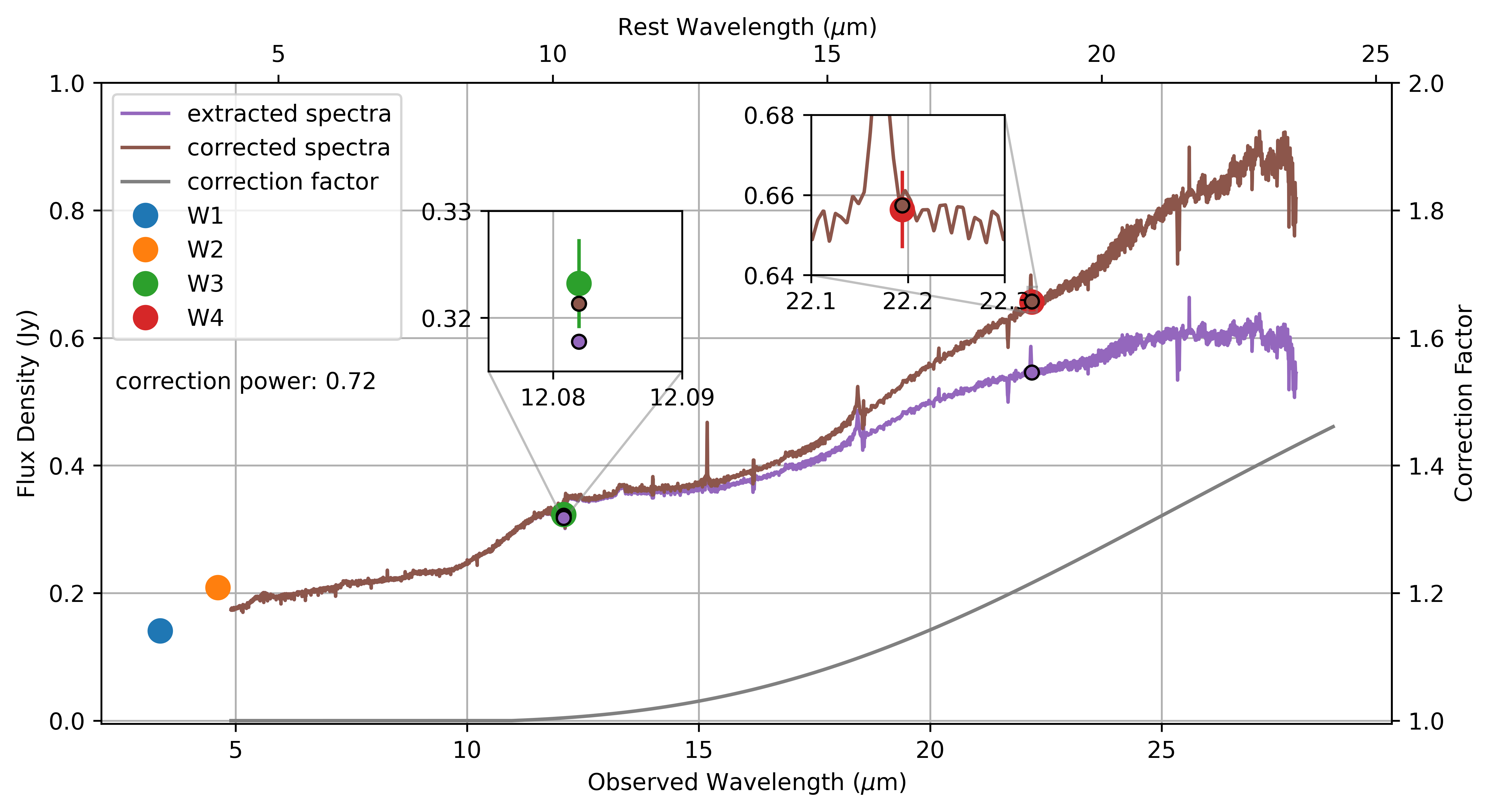}
    \caption{1D spectrum of PDS 456 extracted from the MIRI data cube showing the sensitivity drop at longer wavelengths and its correction. We show the uncorrected spectrum extracted with a conical aperture (purple), the simulated WISE measurements from the \jwst\ spectrum (purple points), the correction factor derived from the \jwst\ correction (gray), the \jwst\ spectrum corrected with this factor (brown), the simulated WISE measurements from the corrected \jwst\ spectrum (brown points), and the WISE measurements (colored points). The left axis displays the flux density in Jy for the spectrum and the right axis displays the scale of the correction factor. }
    \label{fig:1d_corr}
\end{figure*}

For the 1D spectral analysis, we extract a nuclear spectrum for PDS 456 using a cylindrical+conical aperture with a $FWHM(\lambda) = 0.7$\arcsec\ for $\lambda < 8$ \um\ and $FWHM(\lambda) = 0.7 \times \lambda[$\um$]/8$\arcsec\ for $\lambda \geq 8$ \um. This is a very similar method to \cite{Bos2023}, but the initial aperture size was increased to prevent jumps between channels and to scale our data up to the Wide-field Infrared Survey Explorer (WISE) W3 band measurement. Data from each channel are then concatenated to form a longer spectrum through interpolation of overlapping regions. We have rebinned the spectrum on the grid with $\Delta\lambda$ = 8 \AA, the spectral element width of Channel 1. This raw extraction can be seen as the purple spectrum in Figure \ref{fig:1d_corr}. 

In this extracted spectrum, we see a offset between the WISE measurements simulated from the \jwst\ spectrum (purple points) and the colored WISE measurements at the longest wavelengths. This is most clear in W4 (red) which has a $20\%$ offset, but also seen slightly in W3 (green) which has a $2\%$ offset. We believe this offset is primarily due to a sensitivity drop at the longer \jwst\ wavelengths noted by the \jwst\ MIRI MRS team\footnote{See \href{https://www.stsci.edu/contents/news/jwst/2023/performance-monitoring-reveals-a-decrease-in-the-miri-mrs-throughput-at-the-longest-wavelengths?page=5&filterUUID=0655d914-43ee-4d09-a91b-9f45be575098}{April 21, 2023 \jwst\ Observer News Article}}. This drop is time- and wavelength-dependent and a correction was applied in \jwst\ pipeline v1.11.0\footnote{See \href{https://www.stsci.edu/contents/news/jwst/2023/miri-mrs-data-processing-improvements-are-now-available?page=4&filterUUID=0655d914-43ee-4d09-a91b-9f45be575098}{July 14, 2023 \jwst\ Observer News Article}}, but it does not fully correct for the observed offset between \jwst\ and WISE data.

To mitigate this error and bring the \jwst\ spectrum in line with the WISE fluxes, we take a smoothed version of the correction factor applied by the \jwst\ pipeline (gray profile in Figure \ref{fig:1d_corr}), and then scale it with a power law and multiply it into the raw data until the simulated \jwst-WISE and WISE measurements lie within one percent of each other. We also find similar offsets between \jwst\ and WISE data of two other sources: F2M1106 (PID 1335, PI Wylezalek, Co-I Veilleux) and the Phoenix Cluster (PID 2439, PI McDonald). The simulated \jwst-WISE points are lower by $15\%$ and $21\%$ for F2M1106 and the Phoenix Cluster, respectively.

Although the MIRI data reduction has been improving, it is far from perfect. In addition to the aforementioned sensitivity drop, a few other issues remain which we discuss below. 

The spaxel-by-spaxel photometric calibration of the pipeline is inconsistent across MIRI bands (Short, Medium, Long) over different spatial regions of the IFS cube. When bands are combined into channels this results in jumps of up to 40\% in the continuum at some locations at the boundaries between bands. This can be minimized, but not eliminated, by averaging with nearby spaxels, which reduces the spatial resolution. The problem is relatively unimportant in the analysis of spectral lines that fall in the middle of bands but can render analysis of lines along band boundaries difficult and unreliable. The spatial dependence of this error can also make velocity maps unreliable as some regions are more affected than others, causing false results in those regions. 

Spatial undersampling, an expected side effect of maximizing field of view and spectral resolution of \jwst\ \citep{Law2023}, is another major issue with these data. The side effect in our data is wavelength-independent fringes up to 30\% of the continuum. These fringes are most present in the nuclear region of Channel 3 but also occur at a smaller amplitude in Channel 1, and sparsely in Channel 2. The recommended solution to this issue is to extract spectra with a width similar to that of the PSF and use the extended 4-point dither pattern, the latter of which we have done. While a spectral extraction of that size does solve the fringing issue, as seen in the spectrum in Figure \ref{fig:1d_corr}
it does severely degrade the effective spatial resolution of the data which is key to analyzing the host-galaxy and AGN outflow emission. 

We see additional fringing with amplitude $\sim$ 10-30\% of the continuum across all channels. This is a known MIRI MRS issue and should be corrected by using the residual fringe step of the MIRI pipeline but it is not in the case of PDS 456, likely due to its bright undersampled PSF. We see evidence of the known 12.2 \um\ spectral leak and another wide emission feature at $\sim$ 5.5 \um\ which is unexplained by the AGN emission of PDS 456 and we assume is likely a reduction issue. We also see cosmic-ray showers across all MIRI channels but they do not generally interfere with the key spectral features in our data. 

\subsection{\jwst\ NIRSpec}
\label{subsec:nirspec}
As part of the same program PID 2547\footnote{All of the \jwst\ NIRSpec and MIRI/MRS data used in this paper can be found in MAST: \dataset[10.17909/edxc-ef24]{http://dx.doi.org/10.17909/edxc-ef24}.}, PDS 456 was also observed on March 10, 2023 with the IFS mode of NIRSpec \citep{Bok2022, Jak2022}. We used the disperser-filter combination of G235H-F170LP to cover the wavelength range of 1.65 $-$ 3.15 \um\ with a spectral resolution of 8.7 \AA\ corresponding to 85-150 \kms. The field of view of the NIRSpec instrument is 3\arcsec\ $\times$ 3\arcsec, however, due to our four-point extended dither pattern to improve the PSF sampling the field of view was slightly increased and is closer to 4\arcsec\ $\times$ 4\arcsec. The footprint of the NIRSpec detector over the source is shown in Figure \ref{fig:footprints}. The total integration time was 140 minutes. 

The NIRSpec data of PDS 456 was reduced following the same methods as \cite{Vay2023} and \cite{Vei2023}. Briefly, the data reduction was done with the \jwst\ Calibration pipeline version 1.8.4 \citep{Bus2022} using CRDS version 11.16.16 and context file \jwst\ 1019.pmap. Standard steps were taken in \texttt{Detector1Pipeline}. In \texttt{Spec2Pipeline} imprint subtraction was skipped and extra care was taken to flag pixels affected by open MSA shutters. \texttt{Spec3Pipeline} was skipped altogether due to issues at the time. Instead, a script based on \texttt{reproject} \citep{Rob2023} was used to combine each of the dither positions into a single cube. This custom final step produces a cube with a 0\farcs05 spatial grid and 3.96 $\times$ $10^{-4}$ \um\ spectral sampling. 

\subsection{MUSE}
\label{subsec:muse}

PDS 456 was observed with Multi Unit Spectroscopic Explorer (MUSE; \cite{Bac2010, Bac2014}) instrument on the Very Large Telescope (VLT), operated by the European Southern Observatory (ESO) on June 6, 2019 (PID 0103.B-0767, PI Piconcelli). We are using data from the Wide Field Mode (WFM) of the instrument without adaptive optics. This covers a field of view of 1' x 1' with a spatial sampling of 0.2\arcsec\ x 0.2\arcsec. For our analysis, we crop the data down to 14\arcsec\ x 19\arcsec\ giving more room to the south where there are known companions. This encompasses all of the emission from the quasar and host galaxy and is more manageable computationally for our subsequent PSF decomposition analysis. MUSE covers a spectral range of 0.465 $-$ 0.935 \um\ with a spectral resolution of 1.25 \AA\ corresponding to 40-80 \kms. We access these data from the publicly available ESO Archive Science Portal where the data have been pre-processed by the MUSE pipeline \citep{Wei2020}. 

\section{Data Analysis}
\label{sec:analysis}
\subsection{Integral Field Spectroscopy Fitting}
\label{subsec:IFSfits}
We use the software package \qtdfit\ for the majority of our analysis \citep{Rup2023}. This package is based on the IDL software \ifsfit\ \citep{Rup2014, Rup2017} designed to remove the PSF caused by the central compact quasar emission allowing us to see the much fainter emission from the host galaxy without contamination from the bright PSF. This tool is ideally suited for the analysis of IFS quasar data as, without removing a dominant PSF, the compact nuclear region overwhelms emission from extended regions of the galaxy. More details on the \qtdfit\ software and its use in the analysis of \jwst\ IFS data can be found in \citet{Vei2023, Rup2023a, Vay2023}. 

\qtdfit\ extracts a spectrum to use as a quasar template from either the brightest spaxel in the data, a defined radius around that bright spaxel, or a manually set spectrum. In an initial fit, the template, in combination with emission lines and an exponential starlight model, is fit to the data. The template is scaled up or down with a series of exponentials to match the continuum level of different spaxels and remove data that resembles the nuclear spectra. These initial fits may then be used as initial guesses into a second stage of fitting where more detailed models, such as stellar population synthesis, are used to lower the residuals. The process then loops, refitting emission lines and the total sum continuum until certain residual levels are met. Lines profiles are fit with a specified number of Gaussian components to the spectrum with starlight and nuclear emission removed. \qtdfit\ also imposes a significance cut on each component of each line every iteration. If the fit is not significant enough, it will be removed and fit with fewer components or not fit at all. This entire process is done for every spaxel in the data and can be accelerated with multicore processing. Due to a lack of obvious stellar absorption features in the NIRSpec and MIRI data, the use of exponential starlight models was only warranted for the MUSE data.

Clean data is a prerequisite for this process to work. If the extracted template PSF has artifacts such as fringing, jumps, or cosmic rays, then \qtdfit\ will attempt to remove those features from spaxels that do not have them. Most of these issues are caused by fringes because they masquerade as emission lines in width and strength. Even with a perfect quasar template, when individual spaxels have features like fringing, those features will be left after the removal of the nuclear emission, and then fit as extranuclear emission confounding the result. We attempt to minimize these effects by extracting a PSF template from a larger nuclear region, making a custom-smoothed quasar template, and placing velocity limits on line fits. However, some of these artifacts still remain in the \qtdfit\ output maps. We therefore thoroughly examine the results spaxel-by-spaxel and remove bad fits before presentation of the final \qtdfit\ output maps. 

\subsection{Nuclear Spectrum Fitting}
\label{subsec:singleFits}
\qtdfit\ includes some tools for fitting the MIR nuclear spectrum of quasars. These consist of two separate continuum fitting methods. The first of these methods, \texttt{polyfit}, fits a simple, order-specified, polynomial to the continuum with emission lines masked. Then that polynomial is removed from the data and lines are fit to the residual with Gaussians. This is highly effective in localized regions of the spectrum around line profiles and when there is no need or desire to remove the quasar light. In these conditions, the \texttt{polyfit} continuum method can get the most reliable continuum and line fit. 

The second continuum fitting method is called \questfit, based on an IDL software by the same name \citep{Sch2008, Rup2021}. \questfit\ was built to fit mid-IR spectra of galaxies from the \spitzer\ Quasar and ULIRG Evolution Study \citep[QUEST;][]{Vei2009} using polycyclic aromatic hydrocarbon (PAH) and silicate templates, extinction and absorption models, and blackbodies of various temperatures. This method is effective in fitting broad continuum features but less reliable over narrow wavelength regions causing the line fits to be less accurate. 

In using \questfit\ for the fitting of our extracted nuclear spectra, we follow a very similar procedure to that described in \cite{Vei2009} for their fitting of QUEST galaxies. They fit three extincted blackbodies with temperatures that freely vary. These blackbodies do not represent actual dust components but are a good approximation of the MIR continuum in this region. We set initial temperature estimates for the hot, warm, and cool blackbodies of 1350 K, 575 K, and 50 K, respectively, following the three-component description of MIR blackbodies from \cite{Sch2008}. For PAH features we use the noise-free MIR template spectra \#3 and \#4 from \citet{Smi2007}. These spectra fit PAH features with a $\sim$ 0.13 dex variance in PAH 6.2/7.7 \um\ and 7.7/11.3 \um\ ratios. \cite{Vei2009} found that most galaxies fit one template or the other. These templates are based on \spitzer\ spectroscopic data so they are not perfect for the higher spectral resolutions of MIRI/MRS. However, since the PAH features in PDS 456 are very weak, these templates are found to be adequate for this object. For the silicate features, \cite{Sch2008} created a set of narrow line region (NLR) dust emission templates based on data from \citet{Gro2006}. These use a constant density of $10^4$ \eden\ and an ionization parameter, $U = F_*^{ion}/n_H c$ (where $F_*^{ion}$ is the number of ionizing photons, $n_H$ is the hydrogen density, and $c$ is the speed of light), that varies from 10$^{-3}$ to 10$^1$ in steps of 0.3-0.4 dex, resulting in 13 different models. We fit all 13 of these models with \questfit\ letting $U$ run as described above and use $\chi^{2}_{red}$ of the sum fit to determine the best-fit model. 

We implement two methods to derive PAH fluxes. First, we use the best-fit PAH template from \qtdfit\ to extract fluxes from the PAH 6.2, 7.7, and 11.3 \um\ features by simultaneously fitting their main features with a sum of Lorentzian profiles superimposed on a continuum approximated by a second-order polynomial. This follows the methods described in \cite{Sch2006} and should give similar values to those derived in \cite{Rig2021}. Second, we subtract the non-PAH emission of the \questfit\ model from the extracted spectrum and directly fit the residuals with a sum of Lorentzian profiles as in the first method. This method allows for more flexibility in fitting the PAH features, without pre-established constraints on their relative strengths, and thus arguably provides more accurate measurements of the PAH fluxes. However, it is more susceptible to uncertainties in the flux calibration and \questfit\ continuum fitting, reducing its reliability. Both of these fits are shown in Figure \ref{fig:PAHfit}. 

\section{Results}
\label{sec:results}

\subsection{Nuclear Emission}
\label{subsec:nuclear}

In this section, we present the results of fitting the 1D extracted spectrum (Figure \ref{fig:1d_corr}) discussed in Section \ref{subsec:miri}. The overall fit to the spectrum from \qtdfit\ was reasonably good, visually explaining all major expected features of the spectrum, with a $\chi^{2}_{red}$ = 1.86. The broad silicate feature at $\sim$ 10 \um\ is seen in emission, which is common for Type 1 AGN \citep{Hao2005}. The corresponding feature at $\sim$ 18 \um\ is also weakly detected in emission. The best-fit silicate model is \#6 from \cite{Sch2008}. This model has log($U$) = $-$0.6 which corresponds to a incident ionizing flux of log($F_{in}$ [\ergscm]) = 3.96. Like all of the models, it assumes a constant density of $n$(H) = $10^4$ cm$^{-3}$. 

\begin{figure*}
    \centering
    \includegraphics[width=\textwidth]{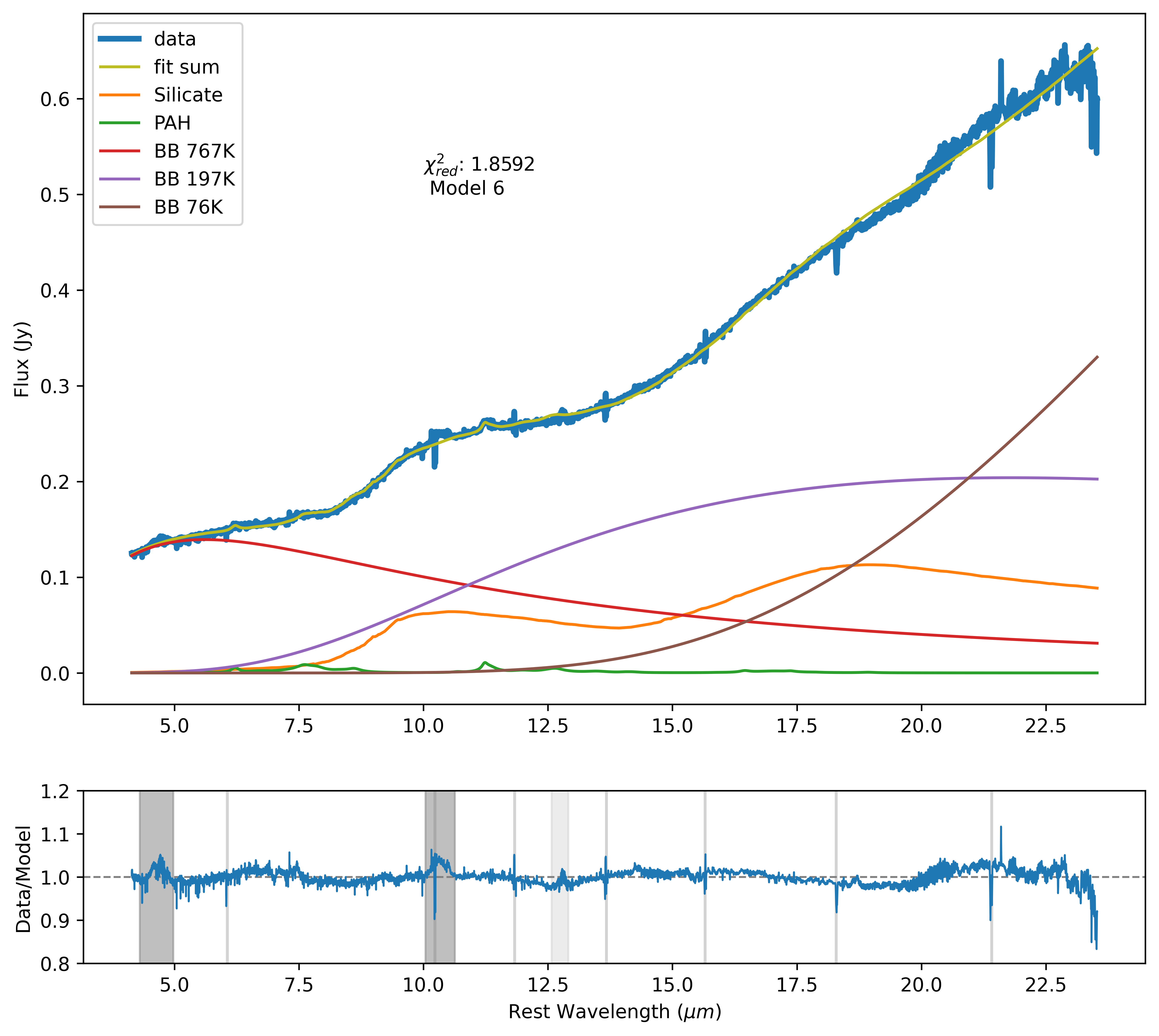}
    \caption{Fit of 1D extracted spectrum (teal) with \questfit\ continuum fitting package within \qtdfit. The spectrum is fit with three black bodies with different temperatures, cool (green), warm (orange), and hot (blue), a PAH template (red), and a silicate emission model (purple). The sum fit is the yellow curve and the residuals of the fit are plotted in the bottom panel. The features in the dark gray boxes on the second axis are data artifacts and the features in the light gray boxes are residual features from the line fitting.}
    \label{fig:1d_fit_new}
\end{figure*} 

Initially, we fit both PAH templates from \cite{Smi2007} to our data to determine which best explains our PAH features. Template \#3 was insignificant in its contribution to the fit of PAH features compared to template \#4, so we fit template \#4 exclusively in our final fit. PAHs are weak in the final fit, but adding PAH template \#4 did measurably improve the $\chi^{2}_{red}$ and visually accounts for features in the PAH regions. Results extracted from our PAH template fit are presented in Table \ref{tab:PAH}.

\begin{deluxetable}{l c c c c}
 \tablecaption{PAH Measurements
 \label{tab:PAH}}
 \tablehead{\colhead{PAH Feature} & \colhead{Method} & \colhead{6.2 \um} & \colhead{7.7 \um} & \colhead{11.3 \um}}
 \startdata
    $\lambda$ range (\um) & & 5.9 $-$ 6.5 & 6.9 $-$ 9.2 & 10.8 $-$ 11.7\\
    EW\tablenotemark{a}& Template & 1.2 (0.2) & 5.9 (0.9) & 2.1 (0.3)\\
     & Direct & 1.4 (0.2) & 5.6 (0.9) & 2.0 (0.3)\\
    Flux\tablenotemark{b} & Template & 2.0 (0.3) & 6.6 (1.0) & 1.7 (0.3)\\
     & Direct & 2.3 (0.3) & 6.2 (0.9) & 1.6 (0.2)\\
    Luminosity\tablenotemark{c} & Template & 5.0 (0.7) & 16.6 (2.5) & 4.3 (0.6)\\
     & Direct & 5.8 (0.7) & 15.7 (2.4) & 4.0 (0.5)\\
 \enddata
 \tablenotetext{a}{In units of $10^{-2}$ \um}
 \tablenotetext{b}{In units of $10^{-13}$ \ergscm}
 \tablenotetext{c}{In units of 10$^9$ $L_\odot$}
 \tablecomments{We derive errors (in parentheses) for our PAH fits by scaling the PAH template up and down until it over- or under-fits the data drastically. Then we calculate the flux values for that scaled template and set them as 1$\sigma$ limits. The method used to fit the nuclear extracted PAH features is listed under the Method column as the template or direct method (described in more detail in \ref{subsec:singleFits}) and illustrated in Figure \ref{fig:PAHfit}.}
\end{deluxetable}

\begin{figure}
    \centering
    \includegraphics[width=.5\textwidth]{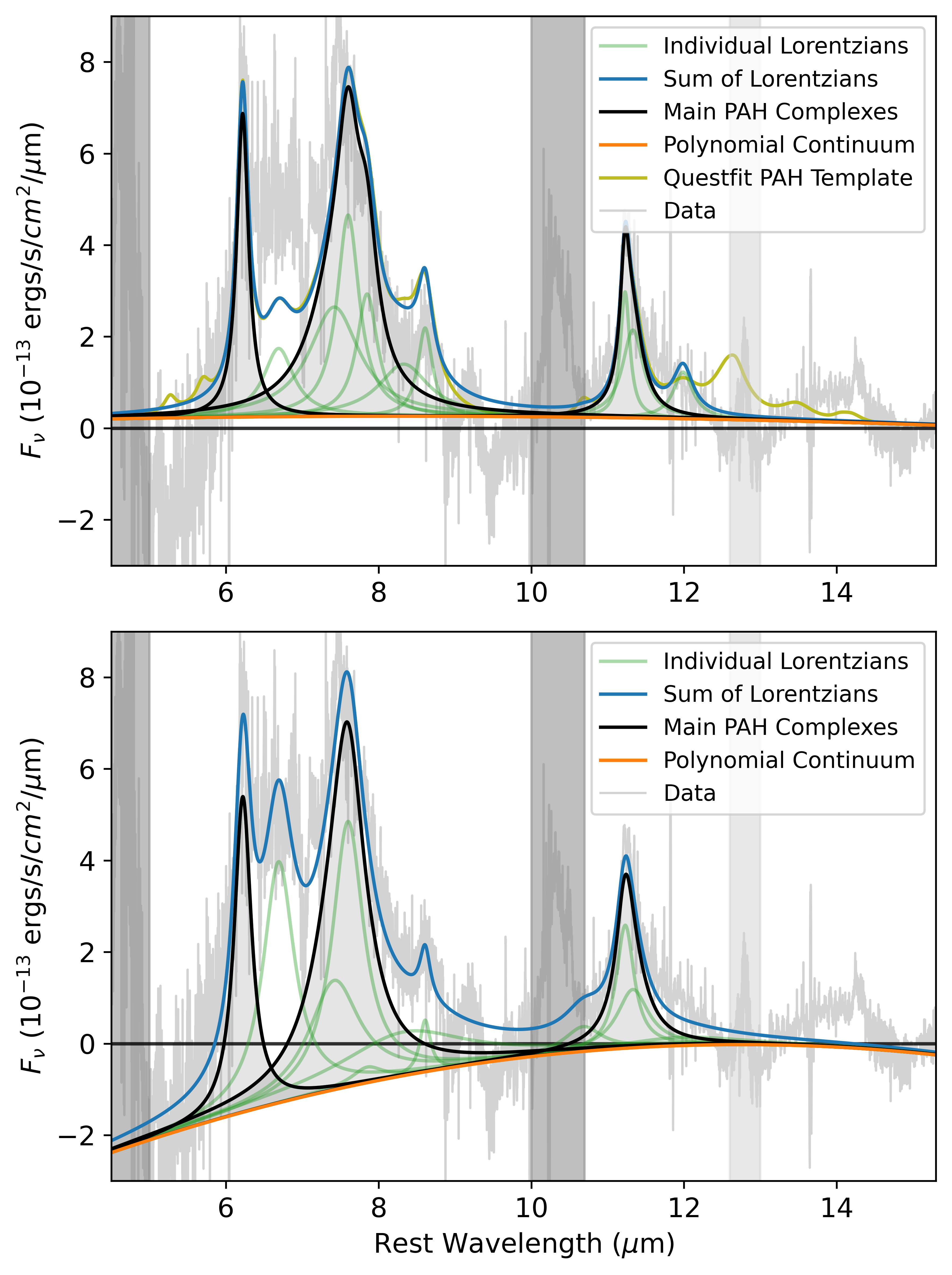}
    \caption{Two methods are used to measure the fluxes of the PAH features in the nuclear spectrum of PDS 456: the \questfit\ template model method (top panel) and the direct fitting method (bottom panel). In the top panel, the  difference between the extracted spectra and the non-PAH emission from the \questfit\ model is shown in gray, the \questfit\ PAH template is shown in yellow-grey, the polynomial fit to the continuum in orange, the individual Lorentzian components to the PAH components in green, the main PAH complexes at 6.2, 7.7, and 11.3 \um\ in black, and the sum of all Lorentzian components in blue. The color code is the same in the bottom panel except there is no \questfit\ template. In this case, the polynomial continuum is allowed to be negative to correct for the imperfect continuum fitting from \questfit. In both panels, the features in the dark gray vertical bands are data artifacts and the feature in the light gray band is a residual from the line fitting around \neii\ 12.8.}
    \label{fig:PAHfit}
\end{figure}

For the fits to the fine structure lines, we first attempt to fit the residuals of the continuum fit from \questfit\ with \qtdfit. However, \questfit\ continuum errors of 2\% near emission lines are not uncommon and can cause errors of 50 times that or more in the fluxes to the fainter lines. This leads us to use the \texttt{polyfit} continuum method instead. We use this method on individual MIRI bands rather than channels or the full MIRI spectrum because the polynomial fit is more reliable over narrower wavelength ranges. We distinctly detect and fit eight emission lines in our data listed in Table \ref{tab:pdslines}. In most cases, a single Gaussian is sufficient to adequately capture the observed line profiles. The only exception is \neiii\ 15.56 \um, where a second Gaussian component is needed to capture a distinct broad blue wing. The core of the other strong lines in the spectrum, from \neii\ and \siii, are much narrower and do not strongly show this blue wing, as shown in Figure \ref{fig:nuc_lines}. Broad, blue wings in higher ionization lines are often seen in AGN-photoionized nuclear outflows \citep[e.g.][]{Vei1991c, Vei1991d, Spo2009, Arm2023}. The resulting fluxes and FWHM (corrected for instrumental resolution effects) from these line fits are listed in Table \ref{tab:pdslines}. 

\begin{deluxetable}{l c c c}
 \tablecaption{Nuclear spectral features
 \label{tab:pdslines}}
 \tablehead{\colhead{Feature ID} & \colhead{$\lambda_{rest}$} & \colhead{Flux} & \colhead{FWHM\tablenotemark{a}} \\
 \colhead{} & \colhead{(\um)} & \colhead{($10^{-15}$ \ergscm)} & \colhead{(\kms)} }
 \startdata
    \niII & 6.636 & 3.03 (1.06) & 490 (120)\\
    \arii & 6.985 & 3.47 (0.63) & 190 (30)\\
    \siv\tablenotemark{b} & 10.511 & 6.83 (1.87) & 1610 (70)\\
    H$_2$ S(2) & 12.279 & 0.61 (0.95) & 80 (90)\\
    \neii & 12.814 & 14.82 (0.77) & 222 (5)\\
    \neiii & 15.555 & 32.21 (4.00) & 790 (30)\\
    H$_2$ S(1) & 17.035 & 2.56 (0.48) & 200 (20)\\
    \siii & 18.713 & 8.03 (0.73) & 350 (10)\\
 \enddata
 \tablenotetext{a}{FWHM of the primary central component. }
 \tablenotetext{b}{\siv\ has considerable contamination due to fringes and continuum variability spaxel to spaxel which is likely causing it to be wider and stronger than expected.}
 \tablecomments{Uncertainties on these fits, listed in parentheses next to the values, are calculated by \qtdfit\ using the residuals from the line fit for the flux and from the fit covariance matrix for the sigma. We use these uncertainties for all lines fit with \qtdfit.}
\end{deluxetable}

\begin{figure}
    \centering
    \includegraphics[width=\columnwidth]{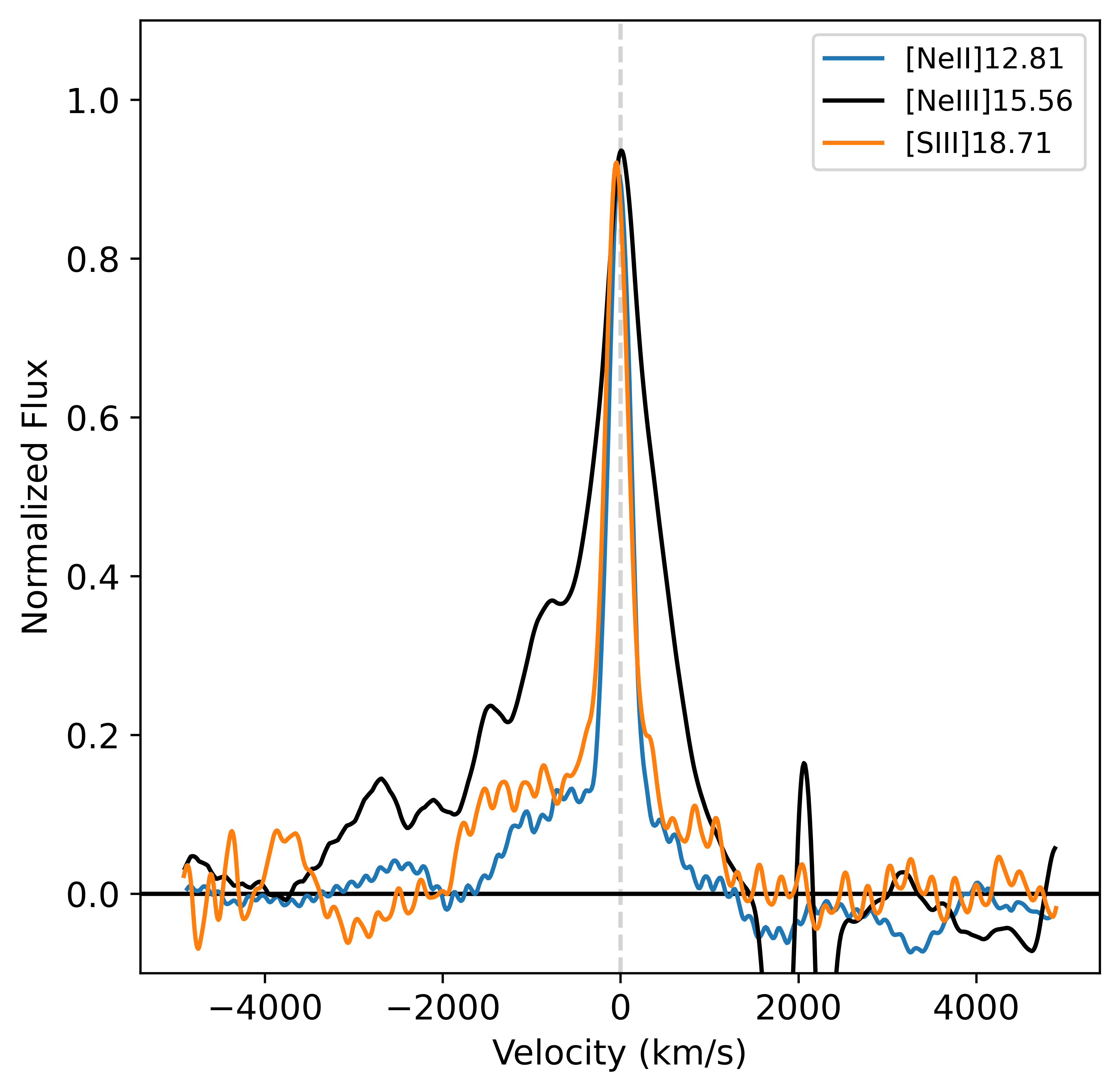}
    \caption{Fine structure atomic lines in the PDS 456 nuclear spectrum. The continuum was fit with a third-order polynomial using \qtdfit\ and the flux values have been normalized to the line peak. The spectra have been lightly smoothed using a moving average to eliminate high frequency fringing.}
    \label{fig:nuc_lines}
\end{figure} 

\subsection{Extranuclear Emission}
\label{subsec:extranuc}

Given the instrumental artifacts in the MIRI data cubes of this source, it is challenging to confidently use these data to ``blindly" search for outflow signatures. Thus, we use the cleaner NIRSpec and MUSE data cubes to help frame and inform our search for outflows in the MIRI data.

\subsubsection{MUSE}
\label{subsubsec:MUSEextended}

A detailed analysis of both the Wide-Field Mode (WFM) and Adaptive Optics Narrow-Field Mode (NFM) MUSE data is presented in \cite{Tra2024}. However, we present our own analysis of the WFM data using \qtdfit\ to allow us to directly compare the results of this analysis with those of the MIRI/MRS data, also based on \qtdfit. We generally find excellent agreement between our results and those from \cite{Tra2024}. 

We run a \qtdfit\ analysis of the cropped MUSE cube across the entire wavelength range, 0.465 $-$ 0.935 \um, so that \qtdfit\ could use the strong \ha\ and \hb\ broad lines from the quasar to assist in the PSF subtraction. To increase the S/N we smooth the cube through a running circular average using a radius of 2.5 spaxels (0.5\arcsec). Across that range, we only fit the \oiii\ 4959, 5007 \AA\ doublet. The results of this fitting are shown in Figure \ref{fig:MUSE}. We show maps of the quasar-subtracted flux, $v_{50}$, and $w_{80}$. $v_{50}$ is the 50-percentile or median velocity: the velocity where 50\% of the line flux is accumulated as calculated from the red side of the profile. $w_{80}$ is the 80-percentile velocity width: the velocity width of the line containing 80\% of line flux, centered on $v_{50}$. See \cite{Vei1991c, Vei1991a, Vei1991b, Zak2014} for a more detailed description and validation of these measurements. 

The $v_{50}$ velocity field primarily highlights the presence of an outflow extending 15 kpc to the east of the center with a median velocity $v_{50}\sim -200$ \kms. The furthest extent of this outflow is just narrowly encompassed by the MIRI channel with the largest field of view. We conclude that the position and kinematics of this emission are inconsistent with the rotating molecular disk detected in the ALMA data of \citet{Bis2019} which lies along a position angle $\sim$ 25$^\circ$, roughly perpendicular to the $v_{50}$ gradient in Figure \ref{fig:MUSE}. 

\begin{figure*}
\centering
\includegraphics[width=\textwidth]{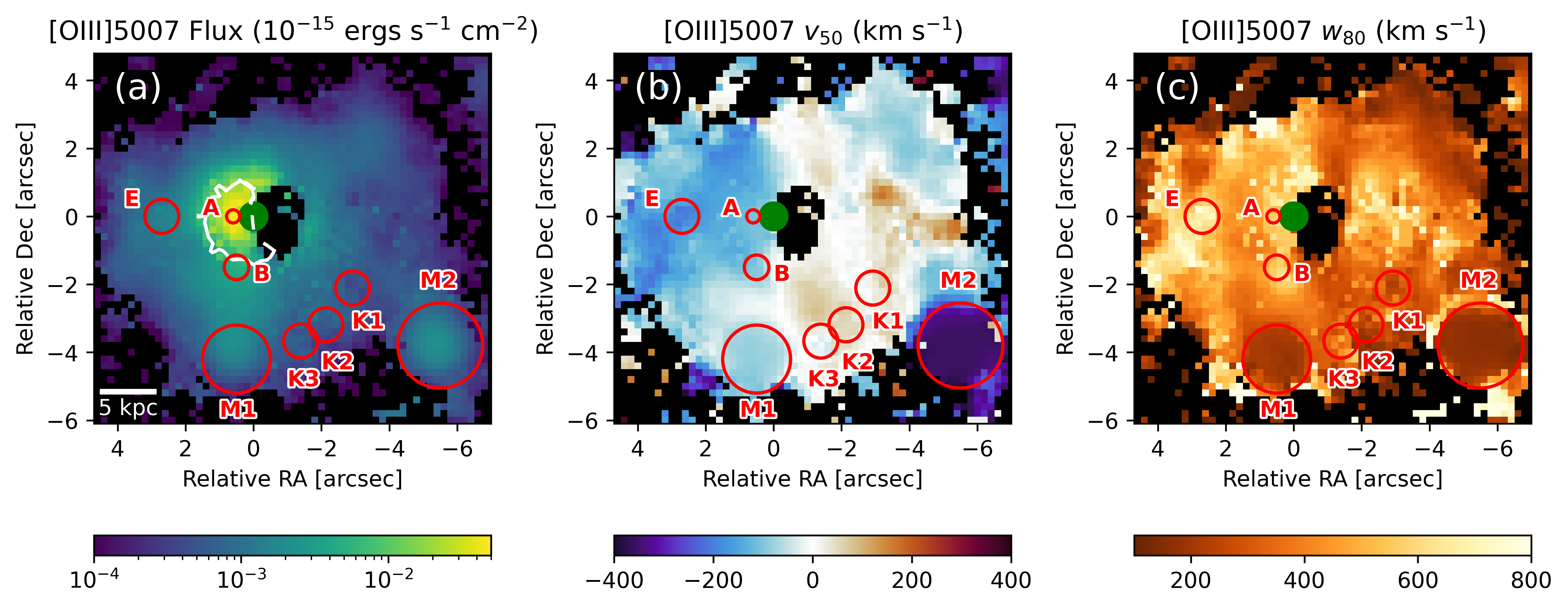}
\centering
\includegraphics[width=.95\textwidth]{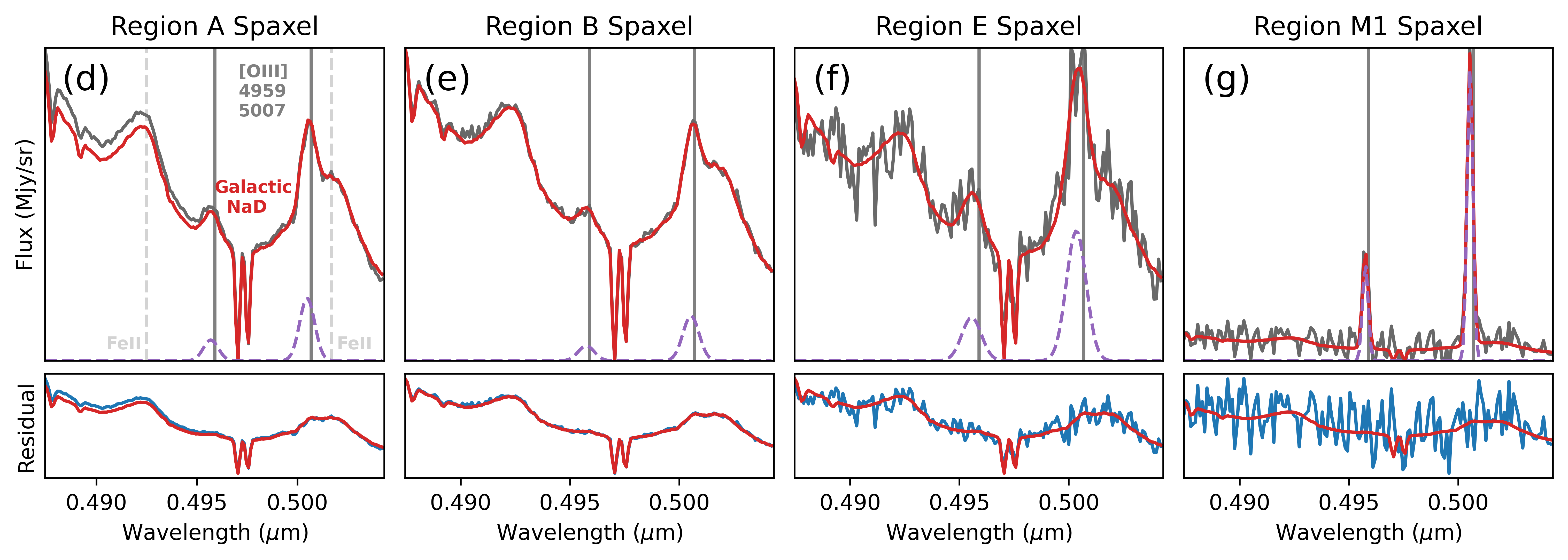}
\caption{\oiii\ flux map, kinematics, and line fits from the MUSE data. (a) Map of the \oiii\ 5007 emission line flux after processing with \qtdfit\ (quasar and stellar continuum emission removed). The white flux contour marks a value of 3 $\times$ $10^{-16}$ \ergscm\ spaxel$^{-1}$. The green dot marks the quasar center with a size of the PSF FWHM.} (b) Map of the 50-percentile (median) velocities, $v_{50}$. (c) Map of the 80-percentile line widths, $w_{80}$. (d)-(g) Representative spectra extracted from the regions indicated in panels (a)-(c). In the top panel of these spectra, the red line is the sum of the scaled quasar template, host continuum model, and the line fit, the last of which is shown as the purple dashed line. The data is the dark gray line. In the bottom panel, the blue spectrum shows the residuals of the data from the line fit and the red line is the sum of the scaled quasar template and host continuum model.
    \label{fig:MUSE}
\end{figure*}

We mark several regions of interest in the MUSE emission maps (Figure \ref{fig:MUSE}). The first set (M1, M2, K1, K2, K3) are companion galaxies. These have been identified by previous studies through continuum and line emission \citep{Tra2024, Bis2019, Yun2004}. The locations and redshifts of M1 (Figure \ref{fig:MUSE}g) and M2 derived from our data are consistent with those from \citet{Tra2024}. Regions K1, K2, and K3 are co-located with strong stellar continuum features fit by \qtdfit, consistent with \citet{Yun2004} and \citet{Tra2024}.

Of the three remaining regions of interest (A, B, and E; Figure \ref{fig:MUSE}d, e, and f, respectively) the first two, regions A and B, spatially correspond to outflowing clumps of CO(3-2) molecular gas from \citet[][corresponding to clumps A and B in their Figure 3a]{Bis2019}. In \oiii\ we do not see the spectrally distinct Gaussians of the core and outflow components as in CO(3-2). However, we do see outflow velocities $v_{90}\sim -450$ \kms\ in line with the molecular outflow velocities from \cite{Bis2019}. Region E has a similar median velocity to regions A and B. 

When comparing our \oiii\ maps to those from \citet[][see Figure 6 in their paper]{Tra2024}, we find largely consistent results. The main differences in our analysis are in the PSF removal method and binning of spaxels to achieve better S/N. They bin spaxels in a 3 $\times$ 3 region whereas we apply a running circular average using a radius of 2.5 spaxels to the cube. Our methods reveal smoother velocity gradients and more distinct substructures in the outflow. This is represented by the redshifted clump to the west of the central emission, which shows up only as a single spaxel in \cite{Tra2024}, and the broader and bluer sub-region near the marked point E, which is non-distinct in \cite{Tra2024}, but is present in other regions.

We note a few issues in our \qtdfit\ analysis of the MUSE data. We believe these are primarily caused by over- or under-fitting the PSF around the \oiii\ line primarily in the central region around the quasar. This causes \qtdfit\ to use the line fit to make up for these fitting errors resulting in extremely broad \oiii\ fits to the east of the quasar and no \oiii\ fits to the west of the quasar. In the region to the east of the quasar strict sigma limits were manually set for each each spaxel based on nearby fits that behaved better. No remediation was taken for the fits in the west region. In both the west and east of the quasar the fits could be fixed by fitting the continuum PSF over a smaller wavelength range around \oiii, but this introduced new issues so the fits were not corrected.

\subsubsection{NIRSpec}
\label{subsubsec:NIRextended}

We run a \qtdfit\ analysis on the first half of the NIRSpec range (1.7 $-$ 2.4 \um) to fit the hydrogen recombination line \paa. We remove the PSF component of \paa\ and fit the residual with a single Gaussian component. A two-component Gaussian fit is appropriate in some regions of the emission but in many other regions, it fits noise or a residual of the PSF which creates noise in the results. Thus we only present the simple single-Gaussian fit for this analysis. To remove noise caused by spectral dips in the brightest spaxels we make a custom quasar template by fitting an extracted (with a circular aperture of radius of 10 spaxels) spectrum with a spline and then removing narrow line emission by fitting a polynomial under the emission. We present our maps of the \paa\ non-nuclear flux, $v_{50}$, and $w_{80}$ in Figure \ref{fig:nirspec}.

Pa$\alpha$ paints a very similar picture to the \oiii\ maps from MUSE. There is a general $v_{50}$ $\sim-200$ \kms\ outflow to the east of the quasar and emission near systemic velocity to the west. The extent of the \paa\ outflow is $\sim 3$ kpc but this is limited by the NIRSpec FOV (Figure \ref{fig:footprints}) and data quality issues at the edge of the detector. 

We mark four regions of interest on the \paa\ maps. Regions A and B correspond to the same locations as those in the MUSE data and regions C and D correspond to notable outflow regions, identified based on \paa\ emission. In region A, we see systemic velocities and a peak in the excess flux emission with $w_{80}\sim400$ \kms. Both regions C and D show higher velocity outflows with $v_{90}\sim-400$ \kms\ and $v_{50}\sim-250$ \kms\ but similar widths as region A. Just to the west of region C (not labelled in Figure \ref{fig:nirspec}), \paa\ has emission profile wider than region C with $w_{80}$ closer to 600 \kms. These larger widths are uncertain due to the complex nature of the \paa\ line. 

In region B, we do not see any significant excess \paa\ emission. There is narrow ($w_{80}\sim150$ \kms) emission at $v_{50}=-480$ \kms\ seen in Figure \ref{fig:nirspec}e which is very similar to the ALMA CO(3-2) emission at region B. However, due to similarities in the line morphology, emission region shape, and strength of the emission to other cosmic ray features, we attribute it to a coincidental cosmic ray instead of ionized outflow. This is supported by a lack of \ha\ emission in the region \citep{Tra2024}. We mark this feature and another cosmic ray close to 1.89 \um\ with the label CR in Figure \ref{fig:nirspec}e. 

We find all of these regions (A, B, C, and D) to be entirely consistent with the \ha\ emission in the MUSE-NFM data cube \citep{Tra2024}. Their analysis shows that the \ha\ excess emission peaks, with systemic velocity, just east of the quasar, spatially consistent with region A in our data. Then to the north, east, and south of this peak, there is a shell of gas with velocities $v_{50}\sim-250$ \kms\ in line with regions C and D. This coincidence between \paa\ and \ha\ emission is expected given their common origin as hydrogen recombination lines (assuming dust extinction does not significantly suppress \ha\ with respect to \paa).

The NIRSpec \paa\ maps also show possible signs of rotation from warm ionized gas (Figure \ref{fig:nirspec}b). There is a north (blueshifted) to south (redshifted) velocity gradient $v_{50, red}\: - \:v_{50, blue} \sim 200$ \kms\ in the data. The general direction and velocities of this gradient are consistent with those of the purported CO(3-2) molecular disk in the ALMA data \citep{Bis2019}. 

\begin{figure*}
\centering
\includegraphics[width=\textwidth]{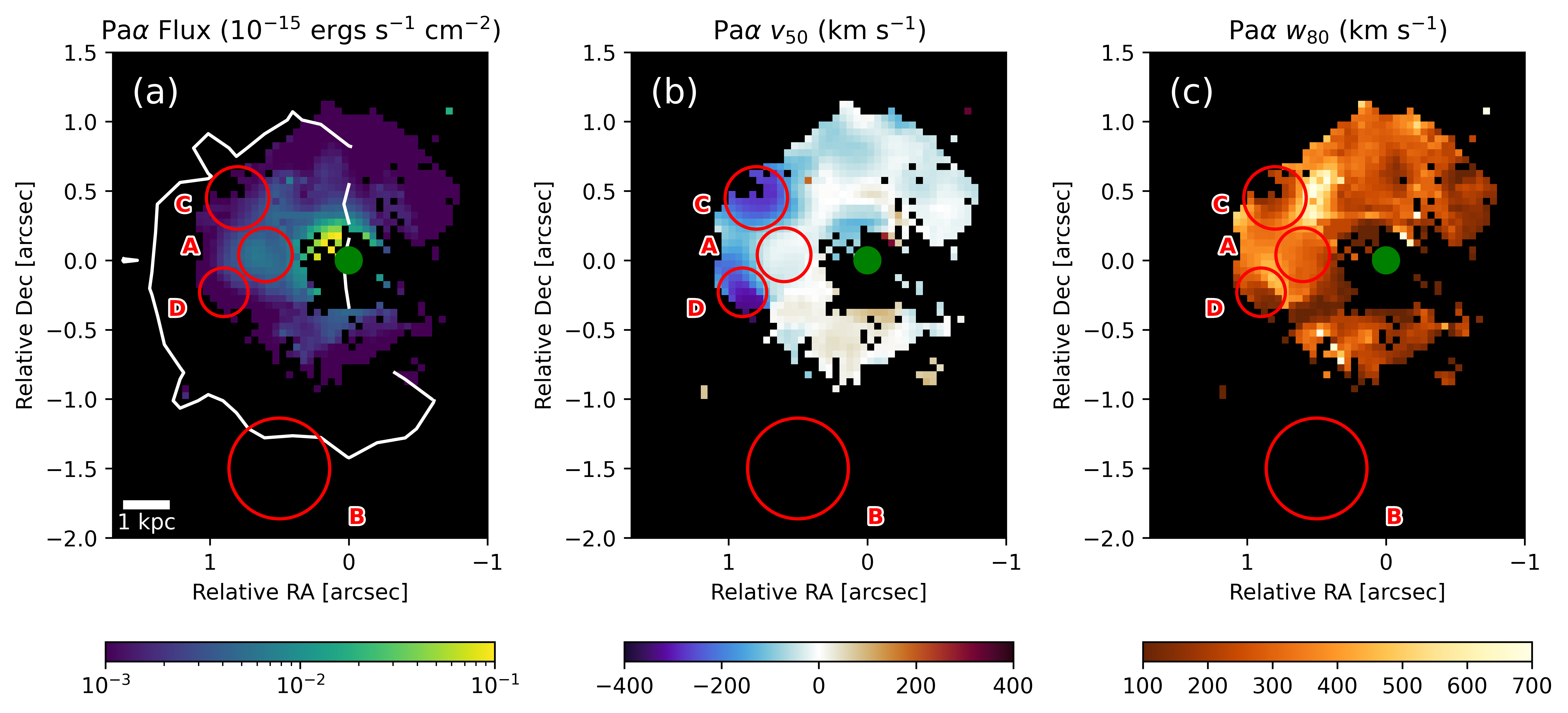}
\centering
\includegraphics[width=.95\textwidth]{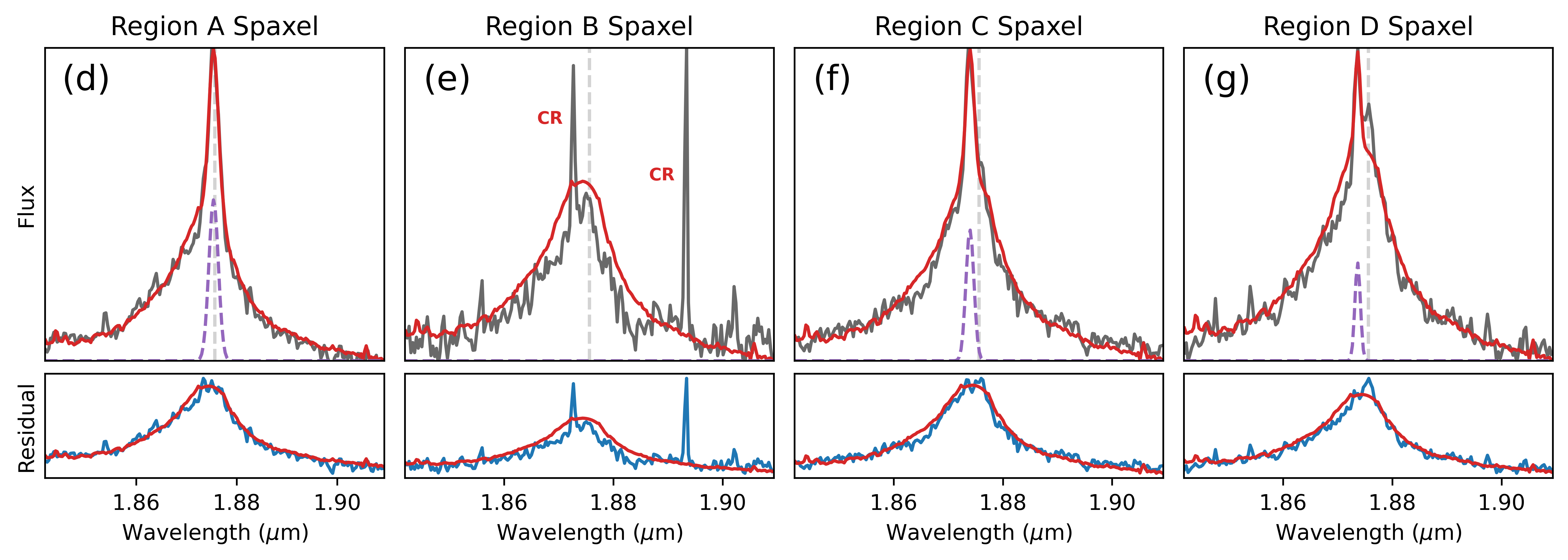}
\caption{\paa\ flux map, kinematics, and line fits from the NIRSpec data. (a) Map of the \paa\ emission line flux after processing with \qtdfit\ (quasar and stellar continuum emission removed). The white contour traces the extended \oiii\ emission shown in Figure \ref{fig:MUSE}a. The green dot marks the quasar center with a size of the PSF FWHM.} (b) Map of the 50-percentile (median) velocities, $v_{50}$. (c) Map of the 80-percentile line widths, $w_{80}$. (d)-(g) Representative spectra extracted from the regions indicated in panels (a)-(c). In the top panel of these spectra, the red line is the sum of the scaled quasar template and the line fit, the latter of which is shown as the purple dashed line. The data is the dark gray line. In the bottom panel, the blue spectrum shows the residuals of the data from the line fit and the red line is the scaled quasar template. Cosmic rays are marked in panel (e) with CR in red.

    \label{fig:nirspec}
\end{figure*}

We run additional \qtdfit\ analysis on the molecular ro-vibrational line transition H$_2$ 1-0 S(3), fitting a narrow wavelength region (1.924 $-$ 1.975 \um) around the line with a \texttt{polyfit} continuum. We find that this fitting method reduces errors, and since H$_2$ 1-0 S(3) emission is not detected in the quasar template there is no need to use the PSF subtraction from \qtdfit. We present our results in Figure \ref{fig:nirspec_h2}. 

The velocity field traced by the H$_2$ 1-0 S(3) emission is relatively consistent with that from \paa. There is outflowing H$_2$ of similar velocity ($v_{50}\sim - 250$ \kms) to the east of the quasar and a cloud near systemic velocity to the west. However, the molecular gas emission is much sparser throughout the whole cube and lacks any emission at region D or to the west of region A. We do not detect H$_2$ 1-0 S(3) emission in region B. Region C to the northeast of the quasar is where most of the blueshifted H$_2$ 1-0 S(3) emission is located and thus is a prime search area for molecular hydrogen lines in the MIRI/MRS data cube. 

The similarities in morphology and location between the warm molecular and ionized gas in NIRSpec imply the presence of both gas phases in the outflow. Our evidence of multiphase outflow in the same dataset supports the findings of \cite{Tra2024} which proposed the co-existence of warm ionized \ha-emitting gas from MUSE and cold molecular CO(3-2)-emitting gas from ALMA in the eastern outflow of PDS 456. 

\begin{figure*}
\centering
\includegraphics[width=\textwidth]{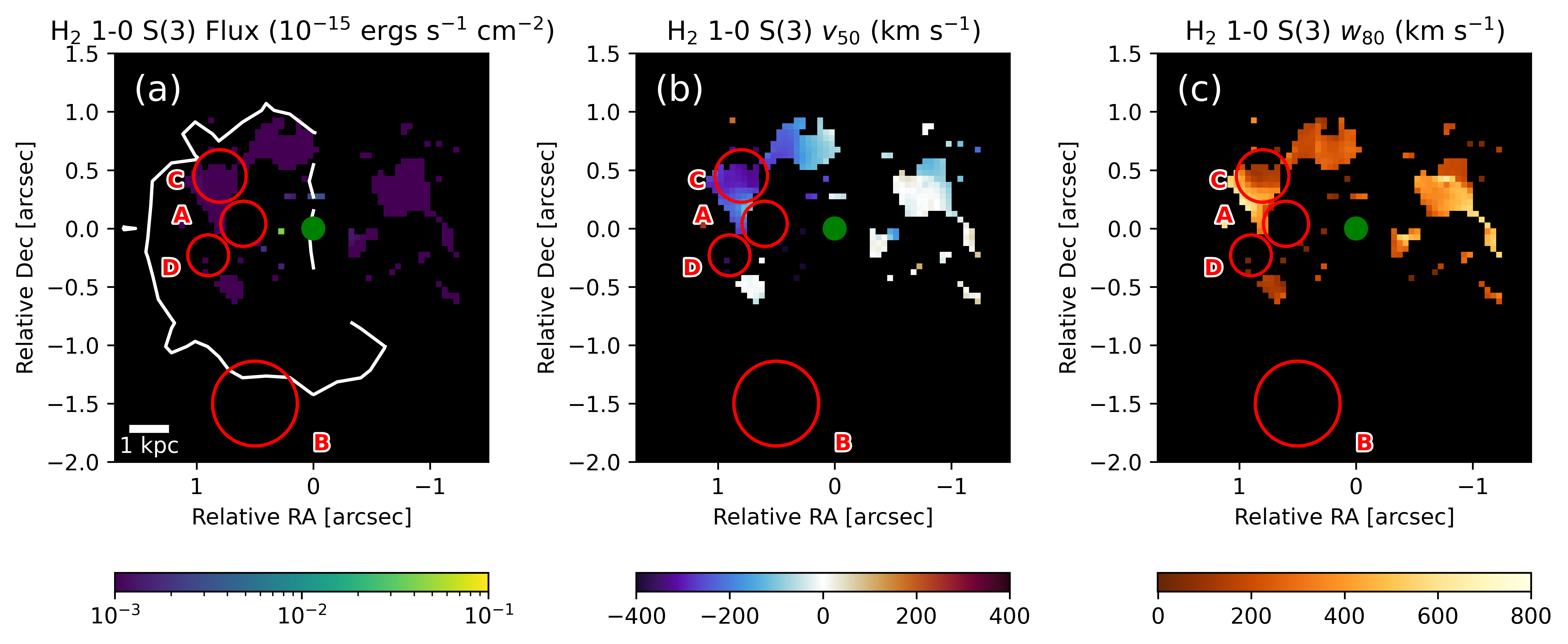}
\caption{H$_2$ 1-0 S(3) flux maps, kinematics, and line fits from the NIRSpec data. (a) Map of the H$_2$ 1-0 S(3) emission line flux. The white contour traces the extended \oiii\ emission shown in Figure \ref{fig:MUSE}a. The green dot marks the quasar center with a size of the PSF FWHM.} (b) Map of the 50-percentile (median) velocities, $v_{50}$. (c) Map of the 80-percentile line widths, $w_{80}$.
    \label{fig:nirspec_h2}
\end{figure*}

Overall, the results from our \qtdfit\ analysis of the NIRSpec and MUSE data paint a picture of a multi-phase outflow to the east of the central quasar, in general agreement with previous results \citep{Bis2019, Tra2024}. We have identified a number of regions of interest where outflowing gas has been detected in the NIRSpec and MUSE data, to help guide our MIRI/MRS analysis. For the warm ionized gas, we expect blueshifted line emission in regions C, D, and E. For the molecular gas, we expect blueshifted line emission in region C.

\subsubsection{MIRI/MRS}
\label{subsubsec:miriextended}

For MIRI/MRS, we fit several emission lines from both the molecular and ionized gas phases (Table \ref{tab:miri_res}). We fit each of the bands where these emission lines are present with \qtdfit. This is done instead of fitting channels or the entire MIRI wavelength range in an attempt to limit PSF removal errors caused by varying photometric calibrations across bands (Section \ref{subsec:miri}). To increase the S/N and limit undersampling, we smooth every band through a running circular average using a radius of 2 spaxels and extract the quasar spectrum for PSF removal with a radius of 1 spaxel, after smoothing. 

In Table \ref{tab:miri_res}, we present an overview of the fit results. For each line, we state if there is nuclear and/or extended emission and if the gas producing this emission is likely outflowing. We also note the local continuum noise level in the data compared to the strength of the line. We did not thoroughly search the entire MIRI converage for other lines not listed in Table \ref{tab:miri_res}. We see a distinct phase gap in lines present in the multiphase outflow. Lower ionization lines like \neii\ and \arii\ are not present in the outflow but higher ionization lines like \neiii\ and \siv\ are. In itself, this is not surprising but we see clear evidence of rotational molecular hydrogen lines in the outflow, which we would expect to be accompanied with an environment also emitting lower ionization fine structure lines. Assuming a multiphase gas, a possible explanation for this evidence is that the molecular gas exists in clumps shielded from the AGN radiation and the dynamic and/or ionization conditions outside the clumps are not favourable for lower ionization line formation.

Of the rotational H$_2$ lines, we see evidence of extended emission up to the S(7) line but no higher. Emission is most clearly near region C in line with the NIRSpec H$_2$ 1-0 S(3) emission. These searches are initially done by eye to search for a detection, then \qtdfit\ is used to confirm that detection in combination with additional visual analysis of the fit. 

\begin{deluxetable*}{l c c c c c c}
 \tablecaption{MIRI/MRS Results
 \label{tab:miri_res}}
 \tablehead{\colhead{Feature ID} & \colhead{$\lambda_{rest}$} & \colhead{IP (eV)} & \colhead{Nuclear} & \colhead{Extended} & \colhead{Outflow} & \colhead{Continuum}\\
 \colhead{} & \colhead{} & \colhead{} & \colhead{Emission} & \colhead{Emission} & \colhead{Signature} & \colhead{Noise Level}\\
  \colhead{(1)} & \colhead{(2)} & \colhead{(3)} & \colhead{(4)} & \colhead{(5)} & \colhead{(6)} & \colhead{(7)}}
 \startdata
    H$_2$ $0-0$ S(9) & 4.69 &  & No & No & No & ---\\
    \feii & 5.06 & 7.9 & No & No & No & ---\\
    \feii & 5.34 & 7.9 & Yes & Yes & Yes\tablenotemark{a} & High\\
    \feviii & 5.45 & 124 & No & No & No & ---\\
    \mgv & 5.61 & 109 & No & No & No & ---\\
    H$_2$ $0-0$ S(7) & 5.51 &  & Yes & Yes & Yes & High\\
    H$_2$ $0-0$ S(6) & 6.11 &  & No & No & No & ---\\
    \niII & 6.64 & 7.6 & Yes & Yes & No & High\\
    H$_2$ $0-0$ S(5) & 6.91 &  & Yes & Yes & Yes & Low\\
    \arii & 6.99 & 15.8 & Yes & No & No  & Low\\
    \naiii & 7.32 & 47.3 & No & No & No & ---\\
    \nevi & 7.65 & 126.2 & Yes & Yes & Yes & Moderate\\
    H$_2$ $0-0$ S(4) & 8.03 &  & Yes\tablenotemark{a} & Yes & Yes & High\\
    \ariii & 8.99 & 27.6 & Yes\tablenotemark{a} & Yes & Yes & High\\
    H$_2$ $0-0$ S(3) & 9.67 &  & No & Yes & Yes & Moderate\\
    \siv & 10.51 & 34.8 & Yes & Yes & Yes & High \\
    H$_2$ $0-0$ S(2) & 12.28 &  & Yes & Yes & Yes & Moderate\\
    Hu$\alpha$ & 12.37 & 13.6 & Yes & Yes & Yes & High\\
    \neii & 12.81 & 21.6 & Yes & Yes & No & Moderate\\
    \nev & 14.32 & 97.1 & Yes & Yes & Yes & High\\
    \neiii & 15.56 & 41.0 & Yes & Yes & Yes & Low\\
    H$_2$ $0-0$ S(1) & 17.04 &  & Yes & Yes & Yes & Moderate\\
    \feii & 17.94 & 7.9 & No & No & No & ---\\
    \siii & 18.71 & 23.3 & Yes & No & No & Moderate\\
 \enddata
 \tablenotetext{a}{Weak detection.}
 \tablecomments{Meaning of the columns: (1) emission line ID, (2) rest wavelength of this emission line, (3) lower ionization potential, (4) is the nuclear emission detected, (5) is the extended emission detected, (6) are there signs of an outflow: extended, blueshifted emission to the east of the quasar with velocities similar to those detected in \paa\ from NIRSpec, (7) continuum noise level in the region surrounding the line, relative to the emission line strength.}
\end{deluxetable*}

We present flux and kinematic maps for five of these lines: three molecular lines, H$_2$ $0 - 0$ S(1), S(3), and S(5), and two fine structure lines, \neiii\ 15.56 and \nevi\ 7.65, tracers of the warm ionized gas. These lines are chosen for their outflow signatures, ionization potentials (IP), and strengths relative to local continuum noise. Spaxels with false positives or bad fits based on a visual inspection of the fits, data, and expected results from the NIRSpec and MUSE analyses were removed from the final maps. The final trimmed maps for the line fluxes, $v_{50}$, and $w_{80}$ are presented in Figure \ref{fig:miri} with regions defined in Sections \ref{subsubsec:MUSEextended} and \ref{subsubsec:NIRextended}.

Spaxels with false positives or bad fits based on a visual inspection of the fits, data, and expected results from the NIRSpec and MUSE analyses were removed from the final maps.

\begin{figure*}
\centering
\includegraphics[width=\textwidth]{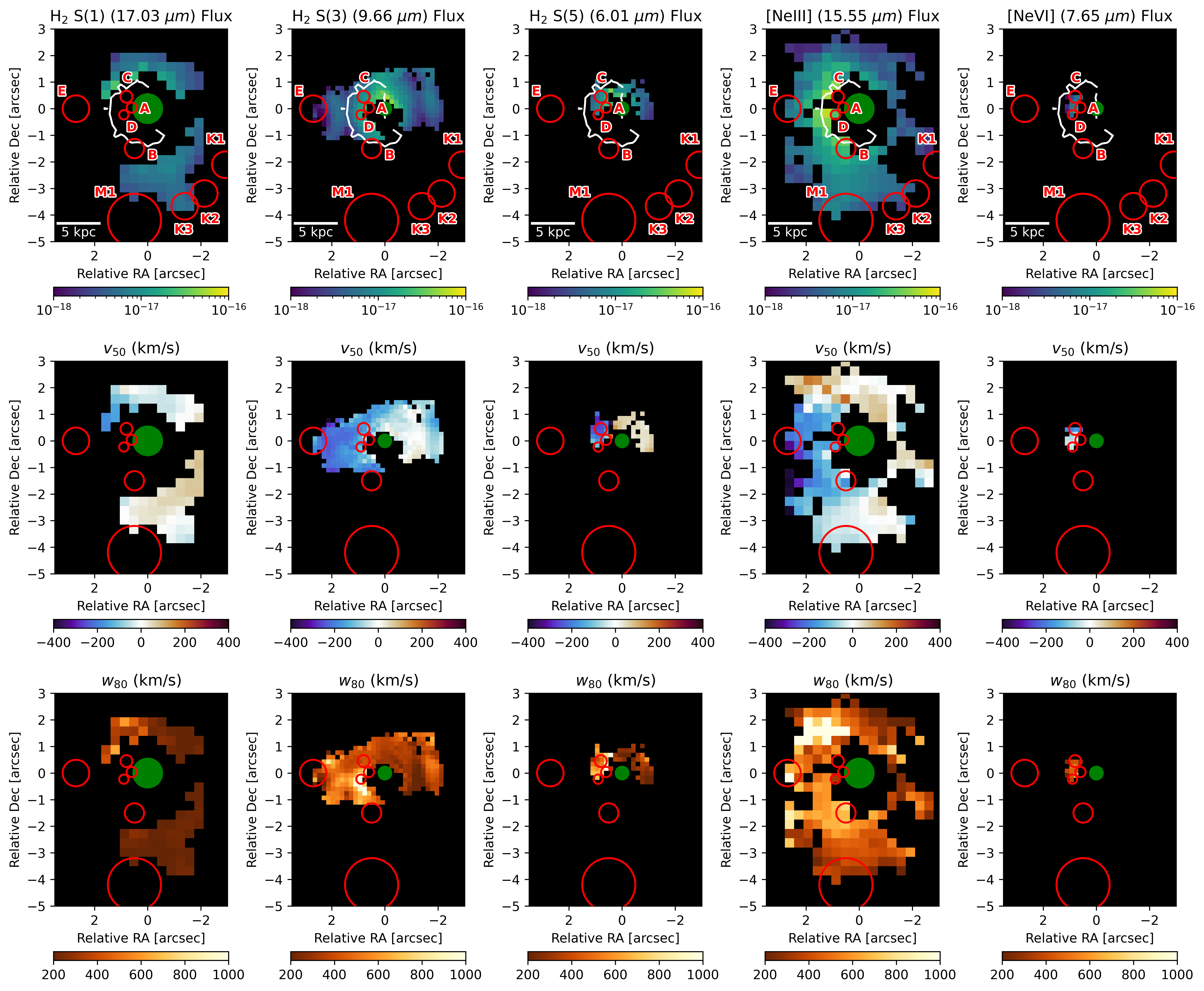}
\caption{Flux and kinematic maps of five lines from the MIRI/MRS data cubes: three molecular lines H$_2$ $0 - 0$ S(1), S(3), and S(5) and two atomic fine structure lines, \neiii\ 15.56 and \nevi\ 7.65. Each column shows maps of the emission line flux, $v_{50}$, and $w_{80}$,  from top to bottom. The white contour in the flux maps traces the extended \oiii\ emission shown in Figure \ref{fig:MUSE}a. The green dot marks the quasar center with a size of the PSF FWHM.} Spaxels with bad fits have been removed from these maps based on a visual inspection informed by the MUSE and NIRSpec fits.
    \label{fig:miri}
\end{figure*}

\begin{figure}
    \centering
    \includegraphics[width=.5\textwidth]{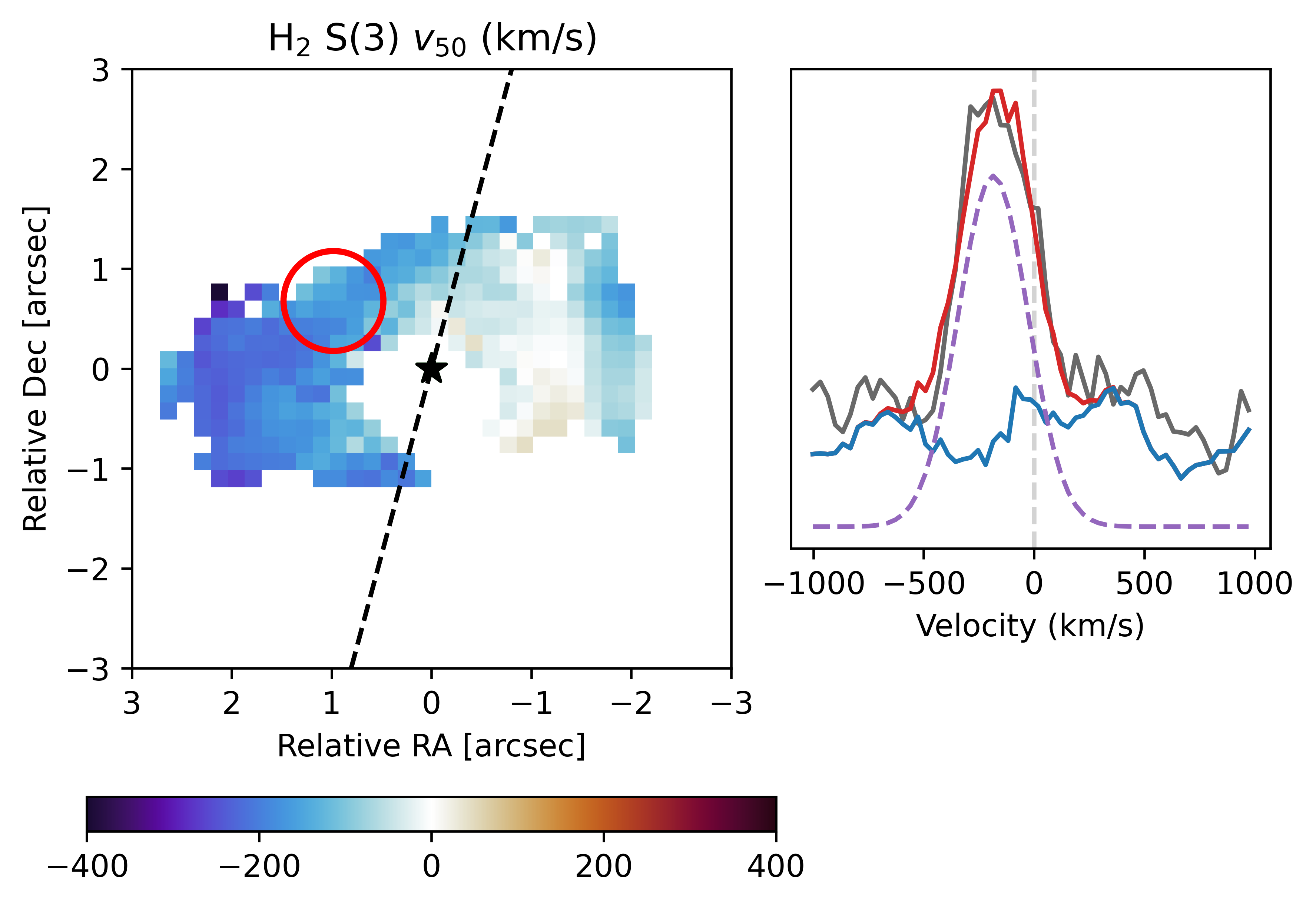}
    \caption{(left) Map of $v_{50}$ for H$_2$ $0-0$ S(3). The kinematic major axis ($PA=165^\circ$, measured from north to east, taken from the \cite{Bis2019} CO(3-2) measurements) is shown as a dashed black line. (right) The gray line shows the  H$_2$ $0-0$ S(3) spectrum extracted from the red circular aperture shown in the left panel, which displays a blueshift $v_{50} \sim -150$ \kms, evidence for the molecular outflow. The red line is the sum of the continuum fit from the scaled quasar spectrum (blue line) and the line fit (purple dashed line).}
    \label{fig:h2s3}
\end{figure}

In these maps, we see the same general trends as in the MUSE and NIRSpec data, with excess blueshifted emission to the east of the quasar and systematic or slightly redshifted emission to the west of the quasar. In the molecular gas phase, H$_2$ 0 $-$ 0 S(3) is the clearest source of extended emission visible in our data set. The maps show a clear velocity gradient in the east-west direction from $v_{90}\sim$ $-$450 \kms\, in the east, to $v_{10}\sim$ +100 \kms\, in the west. The emission extends up to the edge of the MIRI channel 2 detector $\sim$ 8 kpc from the quasar. Velocity widths of the line are maximized directly east of the quasar in region A. An extracted spectrum of this line from a region east of the quasar is presented in Figure \ref{fig:h2s3}. 

The other two H$_2$ rotational lines are generally consistent with this emission. Both the lower state H$_2$ $0 - 0$ S(1) and higher state H$_2$ $0 - 0$ S(5) show blueshifted emission to the northeast of the quasar in the direction of region C. However, both differ in their extent relative to H$_2$ $0 - 0$ S(3). H$_2$ $0 - 0$ S(1) does not have any emission {\em within} the inner 3 kpc and H$_2$ $0 - 0$ S(5) does not have any emission {\em outside} the inner 3 kpc. None of these results is considered physically meaningful: the larger PSF at $\sim$ 20 \um\ is likely responsible for the lack of H$_2$ S(1) emission within the inner 3 kpc (further discussed below), while the faintness of H$_2$ $0 - 0$ S(5) likely explains the lack of detection of this line beyond 3 kpc. 
H$_2$ $0 - 0$ S(5) seemingly shows redshifted emission with higher positive velocities to the west of the quasar, but these velocities are still consistent with those of the other H$_2$ lines given the significant measurement uncertainties in this fainter line. 

In the ionized gas phase, we see extended emission from the highly ionized coronal emission line \nevi\ which has the highest IP in our sample (126 eV). In a small region just east of region A we detect median velocities of $\sim-$100 \kms, slightly slower than that of other emission lines in our sample, but still indicative of outflow given the MIRI data quality. We see no redshifted emission to the west of the quasar in this line. The lower ionization line \neiii\ shows the clearest evidence of resolved ionized outflow. With the unresolved nuclear outflow from the PSF removed there is still clear evidence of extended outflow ($v_{50}$ $\sim-$200 \kms) up to 8 kpc east of the quasar. Other regions to the north and south show more systemic line emission, as expected from the MUSE and NIRSpec results. Median velocities in region M1 are in line with those from MUSE. We note a region with high $w_{80}\sim 1000$ \kms to the north east of the quasar. Close examination of these spaxels reveals that the PSF seems to be underfit and \qtdfit\ is compensating with an overly broad line fit. A smaller $w_{80}\sim 500$ \kms\ is more appropriate to fit the excess \neiii\ emission in this region which is consistent with the kinematics of \oiii\ in the same region (Figure \ref{fig:MUSE}).

The increased size of the PSF at longer wavelengths and significant noise in Channel 4 causes the two lines in this channel (H$_2$ $0 - 0$ S(1) and \neiii) to not show significant excess emission relative to the PSF in the inner 3 kpc from the quasar. 

We further note a region to the southwest of the quasar, near the K3 and K2 companion regions, which shows excess broad, $w_{80}\sim 1500$ \kms, \neii\ and \arii\ emission in addition to strong PAH 11.3 \um. This region is coincident with CO(3-2) line emission in the ALMA data \citep{Bis2019}. 

\section{Discussion}
\label{sec:discussion}

We split our discussion into two parts: nuclear emission and multiphase extended emission. In Section \ref{subsec:agn}, we discuss the results of the nuclear spectrum fitting to calculate a fractional contribution of AGN activity to the bolometric luminosity (Section \ref{subsubsec:agncontribution}), compare our results on the fine structure lines to dusty, radiation pressure AGN dominated photoionization models (Section \ref{subsubsec:agnlines}), calculate the silicate dust cloud distance and covering factor for PDS 456 (Section \ref{subsubsec:silicate}), and use the PAH feature fluxes to calculate SFR and estimate PAH size and ionization (Section \ref{subsubsec:pah}). In Section \ref{subsec:multiphase}, we use our measurements on \neiii\ and the rotational H$_2$ lines to calculate the outflow energetics of the warm ionized gas (Section \ref{subsubsec:ionizeddiscussion}) and warm molecular gas (Section \ref{subsubsec:h2region}), before bringing in the cold molecular gas measurements from \cite{Bis2019} and comparing the results to the central quasar energetics (Section \ref{subsubsec:outflow_overview}). 

\subsection{Nuclear Emission}
\label{subsec:agn}

\subsubsection{AGN Contribution}
\label{subsubsec:agncontribution}
Using the fits of our fine structure lines, PAH features, and continuum measurements described above we adopt the methods from \cite{Vei2009} to calculate the fractional contribution of AGN activity to the bolometric luminosity of PDS 456 (hereafter ``AGN contribution"). \cite{Vei2009} uses their sample of ULIRGs and Palomar-Green quasars to set pure-starburst and pure-AGN limits on AGN contribution and then interpolate between the two and finally convert to the bolometric contribution. With our measurements, we can use four of their six methods. \cite{Vei2009} approximate errors of 10 -- 15 \% on the AGN contribution for all 6 methods. We expect our errors to be at the higher end of this range as we are only using 4 methods. This analysis is done on the extracted nuclear emission. 

The first of these methods uses the equivalent width of the PAH 7.7 \um\ feature. It relies on the underlying assumption that the AGN reduces the PAH equivalent width by contributing additional continuum emission at the PAH wavelength. PAHs may also be destroyed near AGN. Thus, a source with no PAH emission would be considered to be completely AGN-dominated by this method. 

The second method uses a technique originally from \cite{Lau2000}, which was then modified by \cite{Arm2007} and \cite{Vei2009}. The method uses the $f$(PAH 6.2 \um)/$f$(5.3--5.8 \um) and $f$(14--16 \um)/$f$(5.3--5.8 \um) flux ratios with three-component mixing between AGN, H II, and PDR to calculate the AGN contribution.

The third method uses a ratio between MIR and FIR luminosities to calculate the AGN contribution. For the calculation of the MIR luminosity, the emission from the PAHs and silicate is removed from the spectrum fit with \questfit\ and then the remaining flux between 5.4 and 25 \um\ is used to get the luminosity. We calculate the FIR luminosity between 40 and 120 \um\ using the definition of \cite{San1996} and the continuum flux densities from the {\em Infrared Astronomical Satellite} ({\em IRAS}: \citealt{Neu1984})

The fourth method uses the $f_{30}/f_{15}$ continuum ratio as a simpler analog to method \#3. 30 \um\ is beyond the maximum rest wavelength of PDS 456 observed with MIRI (23.5 \um). Thus, we linearly extrapolate our data to get the flux density at 30 \um, a reasonable assumption given the fluxes from {\em IRAS} and {\em ISO}. 

The results of all four methods are listed in Table \ref{tab:agncont}. We get an average AGN contribution to the bolometric luminosity of PDS 456 of 93\% between the four methods. Figure \ref{fig:veilleux} summarizes the results of this analysis and compare the measured values with those of the ULIRGs and quasars from the QUEST sample \citep{Vei2009}. This figure shows PDS 456 solidly in the region dominated by other quasars, well in line with the derived high AGN contribution. The PAH-to-FIR ratio of PDS 456 is typical of other local quasars and Seyfert galaxies. The only outlying feature in this figure is the distinctly higher bolometric luminosity of PDS 456 (Figure \ref{fig:veilleux}c). 

\begin{deluxetable}{l c}
 \tablecaption{Estimates of the AGN Contribution to the Bolometric Luminosity in PDS 456
 \label{tab:agncont}}
 \tablehead{\colhead{Methods} & \colhead{Contribution \%}}
 \startdata
    $W_{eq}$(PAH 7.7 \um) & 77\\  
    Modified \cite{Lau2000} & 99\\
    L(MIR)/L(FIR) & 100\\
    $f_{30}/f_{15}$ continuum ratio & 100\\
    Average & 93\\
 \enddata
 \tablecomments{Column 1: Method used to derive the AGN contribution to the bolometric luminosity, corresponding to methods \#3-6 from \cite{Vei2009}. Column 2: The percentage contribution. The estimated error on the final average value is $\sim$ 10 $-$ 15 \%. }
\end{deluxetable}

\begin{figure*}
    \centering
    \includegraphics[width=\textwidth]{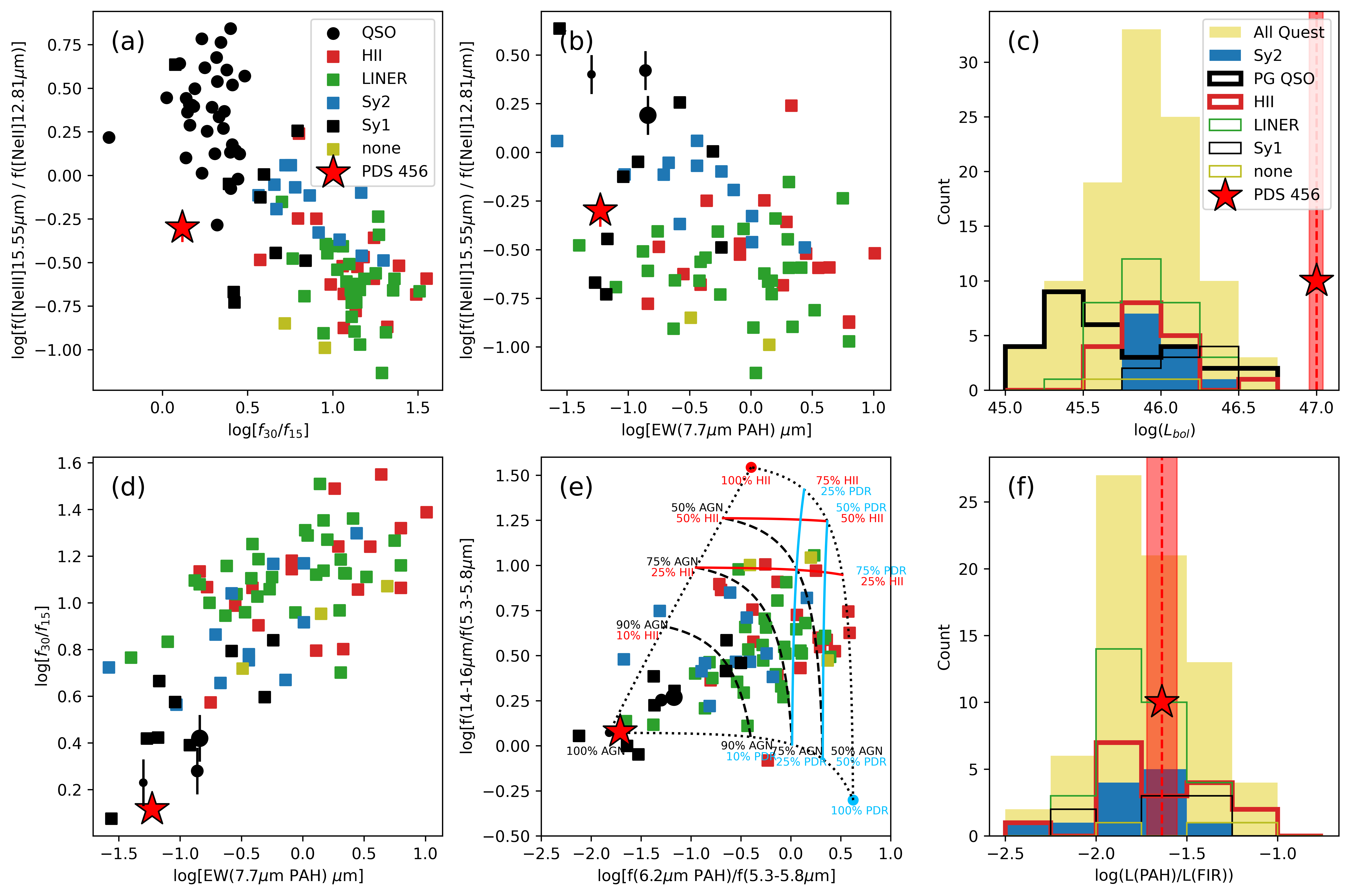}
    \caption{AGN diagnostics adapted from \cite{Vei2009}. (a) \neiii\ 15.56 \um\ to \neii\ 12.81 flux ratios vs.\ the 30-to-15 \um\ flux ratios, (b) \neiii\ 15.56 \um\ to \neii\ 12.81 flux ratios vs.\ the equivalent widths of the 7.7 \um\ PAH feature, (c) L$_{bol}$ histogram of all QUEST sources, (d) Equivalent widths of the 7.7 \um\ PAH feature vs.\ the 30-to-15 \um\ flux ratios, (e) AGN/H II/PDR mixing diagram based on the \cite{Lau2000} method, modified by \cite{Arm2007} and \cite{Vei2009}: PAH (6.2\um) to continuum (5.3--5.8 \um) flux ratios vs.\ the continuum (14--16 \um) to (5.3--5.8 \um) flux ratios, and  (f) 6.2 and 7.7 \um\ PAH luminosities to FIR luminosity ratios. Note that in panels b, d, and e the small, medium, and large black circles correspond to the FIR-undetected, FIR-faint, and FIR-bright Palomar-Green quasars (PG QSOs), as defined in \cite{Net2007}.}
    \label{fig:veilleux}
\end{figure*}

\subsubsection{Fine structure lines}
\label{subsubsec:agnlines}

Since PDS 456 is an AGN-dominated source, we compare our results on the fine structure atomic lines with the predictions from the dusty, radiation pressure-dominated AGN photoionization models from \cite{Gro2004a} to attempt to constrain the ionization parameter and slope of the ionizing power-law spectrum of this source. 
In Figure \ref{fig:lineratios}, we compare the line ratios from our nuclear fits (Table \ref{tab:pdslines}; shown as a red star in each panel of this figure) with the predictions from the solar metallicity models \citep[Table 15 of][]{Gro2004a}. We also consider other metallicities ($Z$ = 0.25, 0.5, 2.0, 4.0 $Z_{\odot}$), but find that changes in the absolute metallicity do not significantly impact the predicted line ratios shown in Figure \ref{fig:lineratios}, as expected given these line ratios largely depend only on relative abundances. To verify the results of our nuclear extraction (which may be affected by instrumental fringing), we also take 0$\farcs$35 radius spectral extractions around regions A and C and measure all of the lines in these spectra, without PSF subtraction. The results are shown as blue and green stars in Figure \ref{fig:lineratios}. While the smaller extraction radius results in larger uncertainties on these measurements, we find that the results are largely consistent with those based on the nuclear extraction.  

\begin{figure*}
    \centering
    \includegraphics[width=\textwidth]{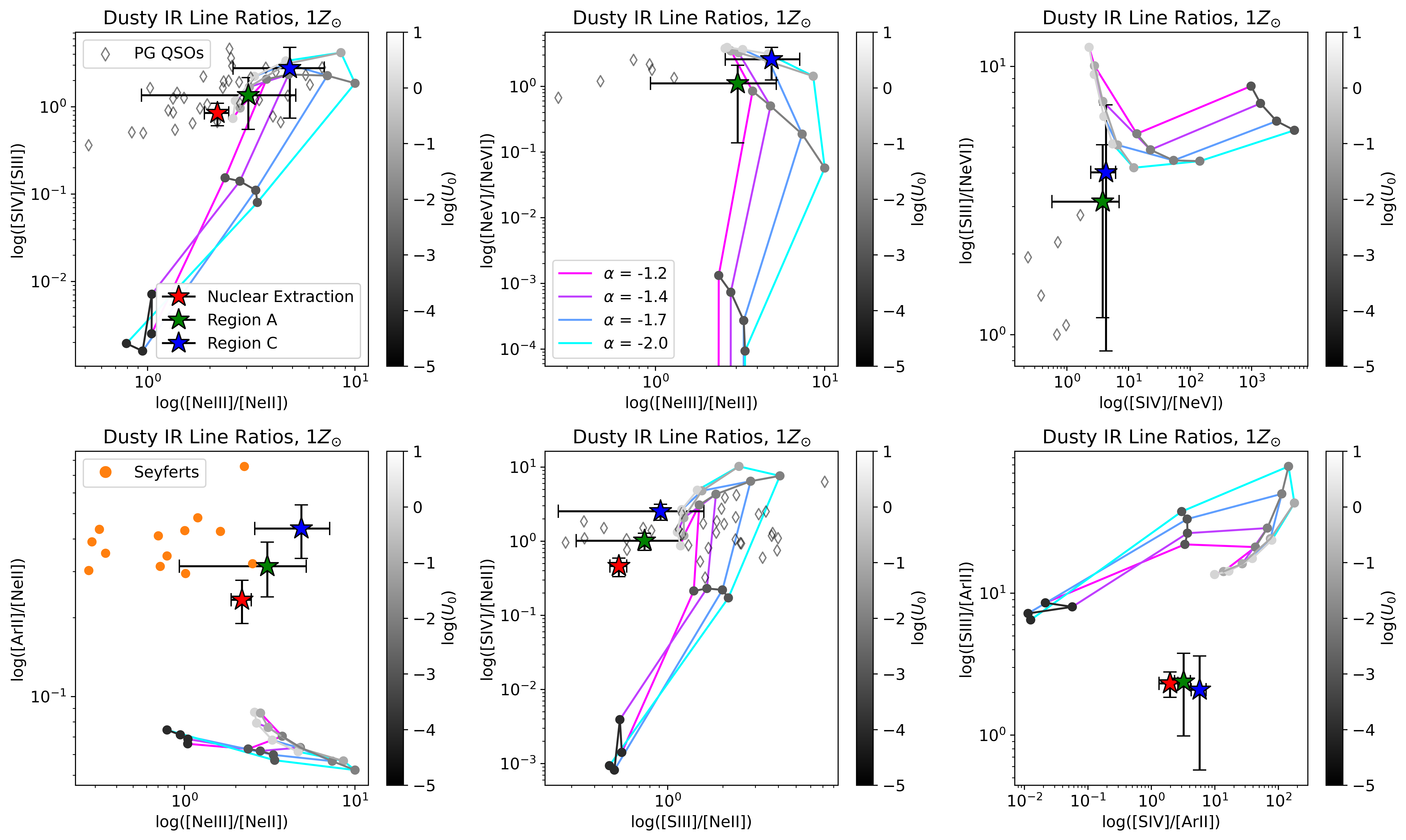}
    \caption{Ratios of the fine structure lines emitted by the warm ionized gas compared to the predictions of dusty photoionized AGN models from \cite{Gro2004a}. We plot the line ratios from three regions: nuclear (red), region A (green), and region C (blue), all without PSF removal. The model grids show the dependence on the ionization parameter ($U$; grayscale nodes) and power-law index of the ionizing source ($\alpha$; pink-cyan lines). The predictions do not depend significantly on metallicity (it is fixed to solar here). We also plot the sample of Seyfert galaxies from \citet[][orange circles]{Gar2022} and the sample of PG QSOs \citet[][gray diamonds]{Vei2009} in order to corroborte our results. See text for more details.}
    \label{fig:lineratios}
\end{figure*}

Overall, the measured line ratios in PDS 456 are reproduced by the radiation pressure-dominated models and are consistent with the sample of Palomar-Green quasars from \cite{Vei2009}. There are, however, two notable exceptions in Figure \ref{fig:lineratios} and both of them involve the [Ar~II] 6.99 line flux; this issue is discussed in the next paragraph. If we ignore these two discrepancies for the time being, the emission line ratios point to a dimensionless ionization parameter, log(U), which lies at the high end of the model range from [$-$1, 0]. Note that the model predictions of \citet{Gro2004a} assume a constant density of $n_H = 10^3$ \eden\ and are not calculated for values of log(U) $>$ 1. For comparison, we derived log(U) = $-$0.6 using the silicate emission (Section \ref{subsec:nuclear}), but this value assumed a higher $n_H = 10^4$ \eden. These two values of log(U) are therefore consistent with each other if the gas emitting the fine structure lines is of lower density (possibly further) than the material emitting the silicate feature. Additionally, our results show a slight preference for AGN models with a shallower ionizing power law of the source: $\alpha = [-1.2, -1.4]$. 

As mentioned earlier, the two clearest discrepancies from the \cite{Gro2004a} models involve the \arii\ 6.99 flux values. The measured \arii/\neii\ and \siii/\arii\ ratios shown in Figure \ref{fig:lineratios} lie +0.2 dex and $-$0.4 dex with respect to the model predictions, respectively. Other line ratios involving \siii\ and \neii\ are consistent with the models. This implies that the models underpredict the observed \arii. This is also true of the \arii/\neii\ values measured in Seyfert galaxies by \citet[][orange circles in the bottom left panel in Figure \ref{fig:lineratios}]{Gar2022}. Thus we conclude that this is unlikely an issue with our data reduction, fitting methods, or a physical explanation unique to PDS 456; instead, it is most likely an issue with the models.

This issue is unlikely to be due to the solar abundances assumed in \citet[][log(Ar/O) = $-$2.29]{Gro2004a}. These are in line with, and even less Ar-depleted, than more recent models \citep[log(Ar/O) = $-$2.36;][]{Nic2017, Are2024}. This is also unlikely to be due to dust depletion since both Ne and Ar are noble gasses and should not interact with dust. We are unable to get good enough S/N on the other Ar lines within the MIRI range such as \ariii\ to determine if other Ar species are also underpredicted by the models. The most likely source of errors in the models is the use of incorrect collision strengths (private discussion with B.\ Groves).

\subsubsection{Silicate Emission}
\label{subsubsec:silicate}

With such luminous AGN, $L_{bol}\sim 10^{47}$ \ergs, it is unsurprising to see silicate in emission due to the significant amount of energy being pumped into the host galaxy which heats the dust. Using Equation 5 from \cite{Sch2008}, we calculate the distance from the central AGN of this emitting dust cloud:
\begin{equation}
\label{eqn:Rdust}
    R_{dust} = \sqrt{\frac{L_{bol}}{4\pi F_{in}}},
\end{equation}
where $F_{in}$ is the incident flux from the NLR model (Section \ref{subsec:nuclear}). With $L_{bol}\sim 10^{47}$ \ergs, we get $R_{dust}$ = 300 pc. As expected, this is much further than the hotter dust measured by \citet{GRA2020} in the near-infrared, which extends to only 1.342 pc. Our dust cloud distance measurement is 40 pc further than the greatest cloud distance derived by \citet{Sch2008} in the Palomar-Green quasars . This is an expected result given the strength of the central engine in PDS 456. Following this, we calculate the NLR covering factor \citep{Sch2008},  

\begin{equation}
\label{eqn:cover}
    c = \frac{F_{\mathrm{NLR}_{\mathrm{fit}}}}{F_{R_{\mathrm{dust}}}}\frac{D^2_L}{R^2_{\mathrm{dust}}},
\end{equation}
where $F_{\mathrm{NLR}_{\mathrm{fit}}}$ is the flux of the silicate model integrated over the observed wavelength range, and $F_{R_{\mathrm{dust}}}$ is the total model flux integrated over the rest-frame wavelength at a distance $R_{\mathrm{dust}}$ from the source. $D_L$ is the luminosity distance to the source. We get $c$ = 0.19, which is typical of the values measured in the Palomar-Green quasars \citep{Sch2008}. 

\subsubsection{PAH}
\label{subsubsec:pah}

The mid-IR PAH features are ubiquitous in galactic and extra-galactic sources. These features are strongly correlated with SFR in normal non-active galaxies \citep{Pee2004, Shi2016, Li2020}. However, AGN suppression of PAH features can affect the SFR-L$_{\mathrm{PAH}}$ relation. This is done through the destruction of PAH molecules \citep{Voi1992, Sei2004}. The 11.3 \um\ PAH feature is arguably the least AGN-suppressed PAH feature in our observations and can fairly reliably indicate SFR for galaxies with AGN \citep{Dia2012, Shi2016}. Thus we use Equation 2 from \cite{Dia2012} to derive SFR = 39 M$_{\odot}$ yr$^{-1}$. This is in agreement with a SFR of 42 M$_{\odot}$ yr$^{-1}$ calculated from the \neii\ 12.81 \um\ line and a SFR of 30 -- 80 M$_{\odot}$ yr$^{-1}$ from rest-frame $\sim$1 mm continuum emission \citep{Bis2019}. We apply other SFR-L$_{\mathrm{PAH}}$ relations using the other PAH features \citep{Far2007, Pop2008, Lut2008, Tre2010, Shi2016}, and those agree with our results within the 1-$\sigma$ uncertainties of those relations, which incorporate both the intrinsic errors and the errors on the SFR estimates, when available.

The ratios between the different PAH features may be a good indicator of the PAH grain size, PAH ionization level, and incident radiation field on those PAHs \citep{Rig2021, Mara2022, Dra2021}. In Figure \ref{fig:pahratio}, we plot the PAH ratios for PDS 456 on an ionization grid from \cite{Rig2021}. This grid was used in \cite{Gar2022} to determine the demographics of PAH features in galaxies with AGN. They determined that AGN PAH signatures are indicative of larger and more neutral PAH populations. This can be seen in Figure \ref{fig:pahratio} where we plot their sample of Seyfert galaxies colored with MIR AGN fraction. We show the measurements from both of our PAH fitting methods (see Figure \ref{fig:PAHfit}) in this figure. The smaller PAH 6.2/7.7 ratio in PDS 456 seems indicative of a smaller dust grain species than the Seyfert galaxies, although the uncertainties on the measurements of PDS 456 are significant. On the other hand, there is no clear evidence that PDS 456 is different from the Seyfert galaxies with respect to PAH ionization and strength of the incident radiation field.

\begin{figure}
    \centering
    \includegraphics[width=\columnwidth]{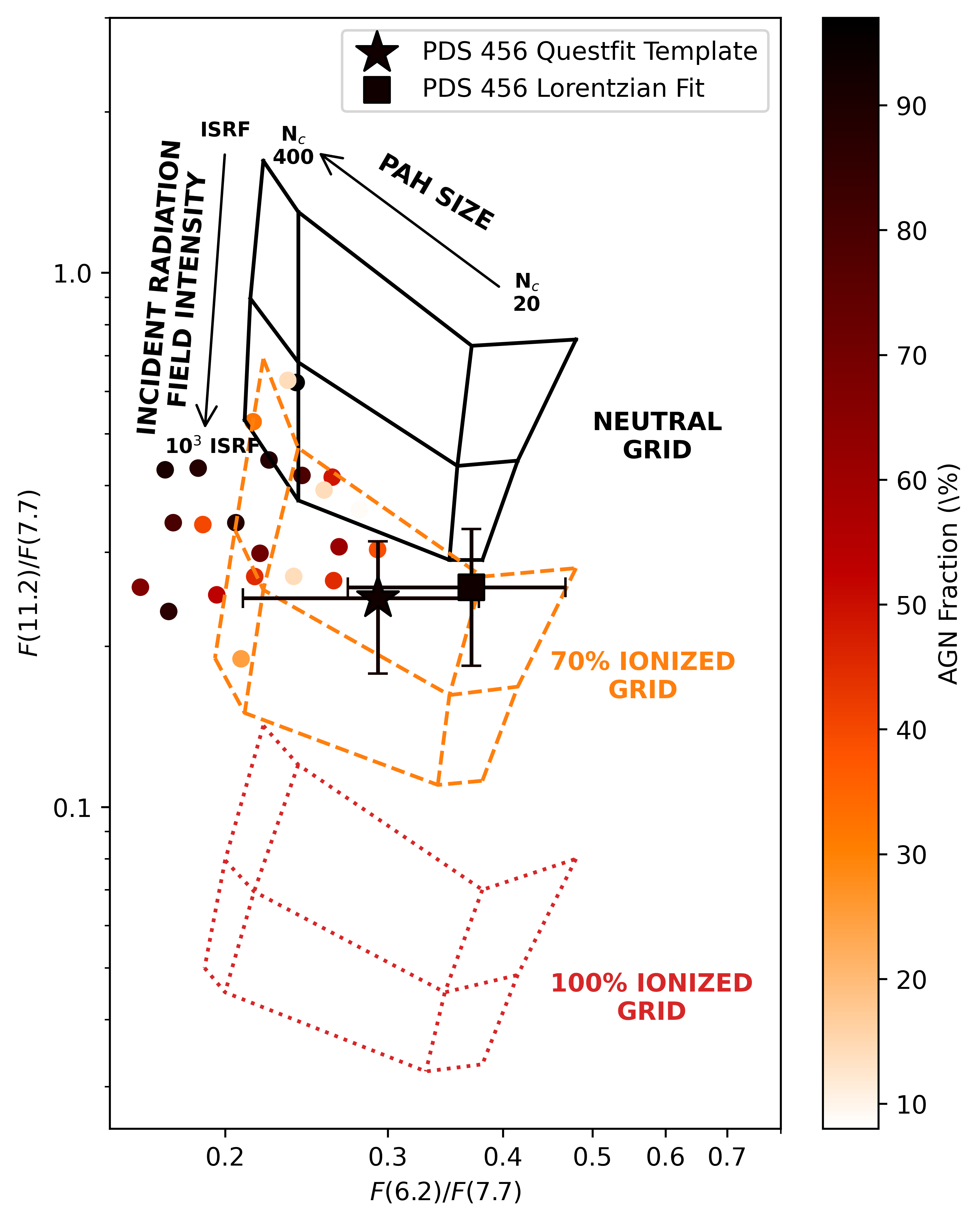}
    \caption{PAH ratio diagnostic diagram. Flux ratios of the PAH 6.2/7.7 \um\ and PAH 11.3/7.7 \um\ from Table \ref{tab:PAH}. The grids are estimated from \cite{Rig2021}. They correspond to mixtures of PAH species ionized to various levels (0\%, 70\%, and 100 \% ionized PAH molecules). The individual grids range from softer (1 $\times$ Interstellar Radiation Field (ISRF); top of the grid) to harder ($10^3$ $\times$ ISRF; bottom of the grid) incident radiation fields and from smaller ($N_c$ = 20, the number of carbon atoms; right side of the grid) to larger ($N_c$ = 400; left side of the grid) PAHs. The colored circles are PAH ratios for Seyfert galaxies colored by their MIR AGN fraction as defined in \cite{Gar2022}. Values for PDS 456 are plotted as the star and square with error bars, and the color reflects the high AGN contribution of PDS 456 (93\%).}
    \label{fig:pahratio}
\end{figure}

\subsection{Multiphase Extended Emission}
\label{subsec:multiphase}

The new MIRI and NIRSpec IFS data allow us to study the spatial and compositional extent of the multi-phase outflow in PDS 456. In this section, we use the measured quantities of the outflowing molecular and ionized gas to derive the physical and dynamical properties of the outflow and discuss the implications. 

\subsubsection{Warm Ionized Gas}
\label{subsubsec:ionizeddiscussion}

We use the luminosities of the outflowing \neiii-emitting gas to estimate the mass of the warm ionized outflow from our data. We choose \neiii\ as it is the strongest fine structure line in our data with both a resolved and unresolved outflow component (seen in Figures \ref{fig:miri} and \ref{fig:nuc_lines}, respectively). To measure the luminosity of the resolved outflow, we take measurements from spaxels directly east of the quasar over a range of about 1.5\arcsec\ to 2.5\arcsec\ in right ascension and $-$1.2\arcsec\ to 1.2\arcsec\ in declination, relative to the quasar. For every spaxel, we take the fit extranuclear component and sum all of the flux with velocities below $-$200 \kms. After summing over all these spaxels,  we then convert the total flux of the outflow into a luminosity using the luminosity distance $D_L$. We choose $-$200 \kms\ as a cutoff because it is well beyond the disk rotational velocity gradient of $-$50 \kms\ to +50 \kms\ for this system \citep{Bis2019}. For the unresolved outflow component, we apply the same method to the fit of \neiii\ from the nuclear spectrum. 

Next, the nuclear and extranuclear line luminosities are used to estimate the mass in warm ionized gas, following a method similar to those of \cite{can2012, Car2015, Vei2020}, except we use the atomic quantities appropriate for \neiii\ 15.5 rather than \oiii\ 5007. Assuming all of the neon is in  Ne$^{2+}$ (reasonable given that there is no clear detection of the outflow in \neii\ and the outflow in \nev\ and \neiv\ is much weaker than in \neiii), $n_\mathrm{H}=10\times n_\mathrm{He}$, a solar neon abundance of [Ne/H] = $-$3.91 \citep{Nic2017}, an emissivity calculated from PyNeb \citep{Lur2015} of $j_{\neiii}=1.93\times 10^{-21}$ \ergs\ cm$^3$ (assuming $T = 10^4$ K which is appropriate for AGN photoionized gas), and a constant electron density $n_e$ to simplify the calculations, we have
 \begin{equation}
 \label{eqn:lneiii}
     L_{\neiii} =  2.8\times 10^{-25} \langle n_e^2 \rangle 10^{\mathrm{[Ne/H]}} f V \mathrm{erg\:s}^{-1},
 \end{equation}
where $\langle n_e^2 \rangle$ is the volume-averaged square electron density, 10$^{\mathrm{[Ne/H]}}$ is the neon-to-oxygen abundance ratio relative to the solar, $f$ is the filling factor, and $V$ is the volume of the outflow. Since the mass of the ionized gas within this same volume is $M_{ionized} =  m_p \langle n_e \rangle f V$, we get
\begin{equation}
\label{eqn:massneiii}
    M_{ionized} = 2.96 \times 10^8 \frac{C\:L_{44}(\neiii) }{\langle n_{e,3} \rangle 10^{\mathrm{[Ne/H]}}} M_{\odot},
\end{equation}
where $C \equiv \langle n_{e} \rangle^2/\langle n_{e}^2 \rangle $ is the electron density clumping factor, which we assume is of order unity, $L_{44}(\neiii)$ is the \neiii\ luminosity normalized to 10$^{44}$ \ergs, $\langle n_{e,3} \rangle$ is the average electron density normalized to 10$^3\:\mathrm{cm}^{-3}$. As mentioned earlier, the mass from this estimate assumes a constant temperature T $\sim10^4$ K,  but note that the temperature dependence is weak. It also assumes that the electron densities are below the critical density, $n_e \lesssim 2.1 \times 10^5\:\mathrm{cm}^{-3}$, so that collisional de-excitation is unimportant. For better comparison to \citet{Fio2017} and \citet{Tra2024} we assume the same electron density value as those studies, $n_e$ = 200 cm$^{-3}$.

\begin{deluxetable*}{l c c c c c c c c}
 \tablecaption{Outflow Parameters
 \label{tab:outlflow}}
 \tablehead{\colhead{Gas Phase} & \colhead{Tracer} & \colhead{Component} & \colhead{$M_{\mathrm{out}}$} & \colhead{$v_{\mathrm{out}}$} & \colhead{R$_\mathrm{out}$} & \colhead{$\dot M_{\mathrm{out}}$} & \colhead{$\dot P_{\mathrm{out}}$} & \colhead{$\dot E_{\mathrm{out}}$}\\
 \colhead{} & \colhead{} & \colhead{} & \colhead{(10$^7\:M_{\odot}$)} & \colhead{(\kms)} & \colhead{(kpc)} & \colhead{($M_\odot \mathrm{yr}^{-1}$)} & \colhead{($10^{33} dyne$)} & \colhead{($10^{40}$ \ergs)}\\
 \colhead{(1)} & \colhead{(2)} & \colhead{(3)} & \colhead{(4)} & \colhead{(5)} & \colhead{(6)} & \colhead{(7)} & \colhead{(8)} & \colhead{(9)}}
 \startdata
    Warm Ionized & \neiii\ 15.56 & Unresolved & 2.4$_{-0.6}^{+0.6}$ & 1000$_{-200}^{+200}$ & 1.5$_{-1}^{+1.5}$ & 15$_{-10}^{+55}$ & 100$_{-40}^{+400}$ & 500$_{-300}^{+2000}$\\
    & & Resolved & 0.04$_{-0.01}^{+0.01}$ & 200$_{-100}^{+100}$ & 6$_{-3}^{+3}$ & 0.01$_{-0.01}^{+0.04}$ & 0.02$_{-0.01}^{+0.08}$ & 0.02$_{-0.01}^{+0.1}$\\
    & & Total & 2.44$_{-0.7}^{+0.7}$ & & & 15$_{-10}^{+35}$ & 100$_{-40}^{+400}$ & 500$_{-300}^{+2000}$\\
    Warm Molecular & H$_2$ 0-0 S(3) & Resolved & 0.027$_{-0.004}^{+0.004}$ & 200$_{-100}^{+100}$ & 5$_{-3}^{+3}$ & 0.01$_{-0.008}^{+0.04}$ & 0.014$_{-0.002}^{+0.08}$ & 0.014$_{-0.006}^{+0.12}$ \\
    Cold Molecular\tablenotemark{a} & CO(3-2) & Total & 25$_{-9}^{+3}$ & & & 290$_{-110}^{+470}$ & 1200$_{-420}^{+2000}$ & 4000$_{-3300}^{+1800}$ \\
 \enddata
 \tablecomments{Meaning of the columns: (1) Gas phase of the outflow, (2) emission line tracer used to derive the mass, (3) component of the outflow, either resolved or unresolved, (4) mass of the outflowing gas phase, (5) typical $v_{50}$ velocity of the outflow,  (6) median radius of the outflow, (7) mass outflow rate,  (8) momentum outflow rate, and (9) energy outflow rate}
 \tablenotetext{a}{Data gathered from Table 1 of \cite{Bis2019}}
\end{deluxetable*}

The calculated masses are listed in Table \ref{tab:outlflow}. These estimates are in very good agreement with those calculated from \oiii, a total (resolved + unresolved) mass of $M = 2.3_{-0.2}^{+0.2} \times 10^7 M_{\odot}$ \citep{Tra2024}. \neiii\ lacks contamination from other nearby lines, unlike \oiii\ which has nearby Fe~II lines. It is also less affected by dust extinction than \oiii. \citet{Tra2024} derive an extinction from \ha/\hb\ of E(B-V) = $0.2-1.1$ mag.\ (lower values further away from the center), which translates to a negligible extinction of $<0.1$ mag.\ in the MIR \citep{Wei2001, Gor2023}. Thus, we consider the \neiii-based mass estimate to be more reliable than the value based on \oiii. 

We use this mass to very roughly estimate the mass outflow rate, 
\begin{equation}
\label{eqn:massoutflow}
    \dot M_\mathrm{out} = \frac{M_\mathrm{out} v_{\mathrm{out}}}{R_\mathrm{out}},
\end{equation}
where $v_{\mathrm{out}}$ is the typical outflow velocity and $R_\mathrm{out}$ is the median distance from the quasar of the outflow.\footnote{Note that this equation is missing a factor of 3 relative to that used to calculate the mass outflow rate in \citet{Tra2024} since we find no strong reason to assume a constant average volume density, which would require a decaying outflow history in this object \citep{Lut2020, Vei2020}.} To calculate $v_{\mathrm{out}}$ and $R_\mathrm{out}$ for the resolved outflow, we take the median velocity, $v_{50}$, and median distance to the central quasar of all spaxels in the outflow region. To calculate $v_{\mathrm{out}}$ for the unresolved outflow, we take the median velocity, $v_{50}$, of the secondary Gaussian fit to the blueshifted wing. For $R_\mathrm{out}$, we take the half distance between the closest resolved outflowing emission and the center of emission. These radii and velocities are listed in Table \ref{tab:outlflow} along with the calculated outflow momentum rate, 
\begin{equation}
\label{eqn:momentumoutflow}
    \dot P_\mathrm{out} = \dot M_\mathrm{out} v_{\mathrm{out}},
\end{equation}
and outflow power, 
\begin{equation}
\label{eqn:energyoutflow}
    \dot E_\mathrm{out} = \frac{1}{2} \dot M_\mathrm{out} v_{\mathrm{out}}^2.
\end{equation}

\subsubsection{Molecular Gas}
\label{subsubsec:h2region}

We see significant excess H$_2$ $v = 0 - 0$ line emission in the outflow in region C, 0$\farcs$66 northeast of PDS 456 (Figure \ref{fig:miri}). We extract a spectrum with a radius of 0$\farcs$35 centered on this region to measure the extensive series of rotational H$_2$ lines present in the data. We use \qtdfit\ to remove the nuclear quasar emission and then fit the lines in this extracted spectrum using a single Gaussian to fit each line. Using emission probablities for the H$_2$ $v = 0 - 0$ transitions from \cite{Rou2019} and Equation 1 of \cite{You2018} to calculate $N(\nu_u, J_u)$, we show, in Figure \ref{fig:temperature}, an excitation diagram for all of the H$_2$ $0-0$ lines present in region C (S(1), S(2), S(3), S(4), S(5), S(7)). The H$_2$ temperature may be inferred from a linear fit to the data using Equation 2 from \cite{You2018}:

\begin{equation}
\label{eqn:h2temp}
    log_{10} \frac{N(v_u, J_u)}{g(J_u)}=-\frac{1}{T\cdot ln(10)} \frac{E(v_u, J_u)}{k_B} + log_{10} N(0,0). 
\end{equation}
In practice, we derive three different temperatures from three different fits using either all the lines, only the lines sensitive to cooler temperatures (S(1), S(2), S(3), S(4); T$_{cool}$), or only the lines sensitive to hotter temperatures (S(4), S(5), S(7); T$_{warm}$). In all cases, we assume an ortho/para ratio of 3 for the statistical weights in the above equation. Smaller ortho/para ratios have been observed in some cases \citep[e.g.][]{Hab2005}, but a value of 3 is a conservative upper limit useful for comparisons. Smaller values would slightly decrease the calculated cooler temperatures, increase the warmer temperatures, and increase the mass estimates (by a factor of 60\%). 

The fits for each group of lines and the derived temperature for each fit are displayed in Figure \ref{fig:temperature}. We get T$_{all}=810\pm140$ K, T$_{cool}=500\pm120$ K, and T$_{warm}=1420\pm260$ K. Since there is no detectable silicate absorption from $\tau_{9.7}$ (no extinction measured by \questfit\ or seen in H$_2$ $0-0$ S(3) relative to the other H$_2$ $0-0$ lines), we do not expect significant extinction, as discussed in Section \ref{subsubsec:ionizeddiscussion}. 

\begin{figure}
\centering
\includegraphics[width=\columnwidth]{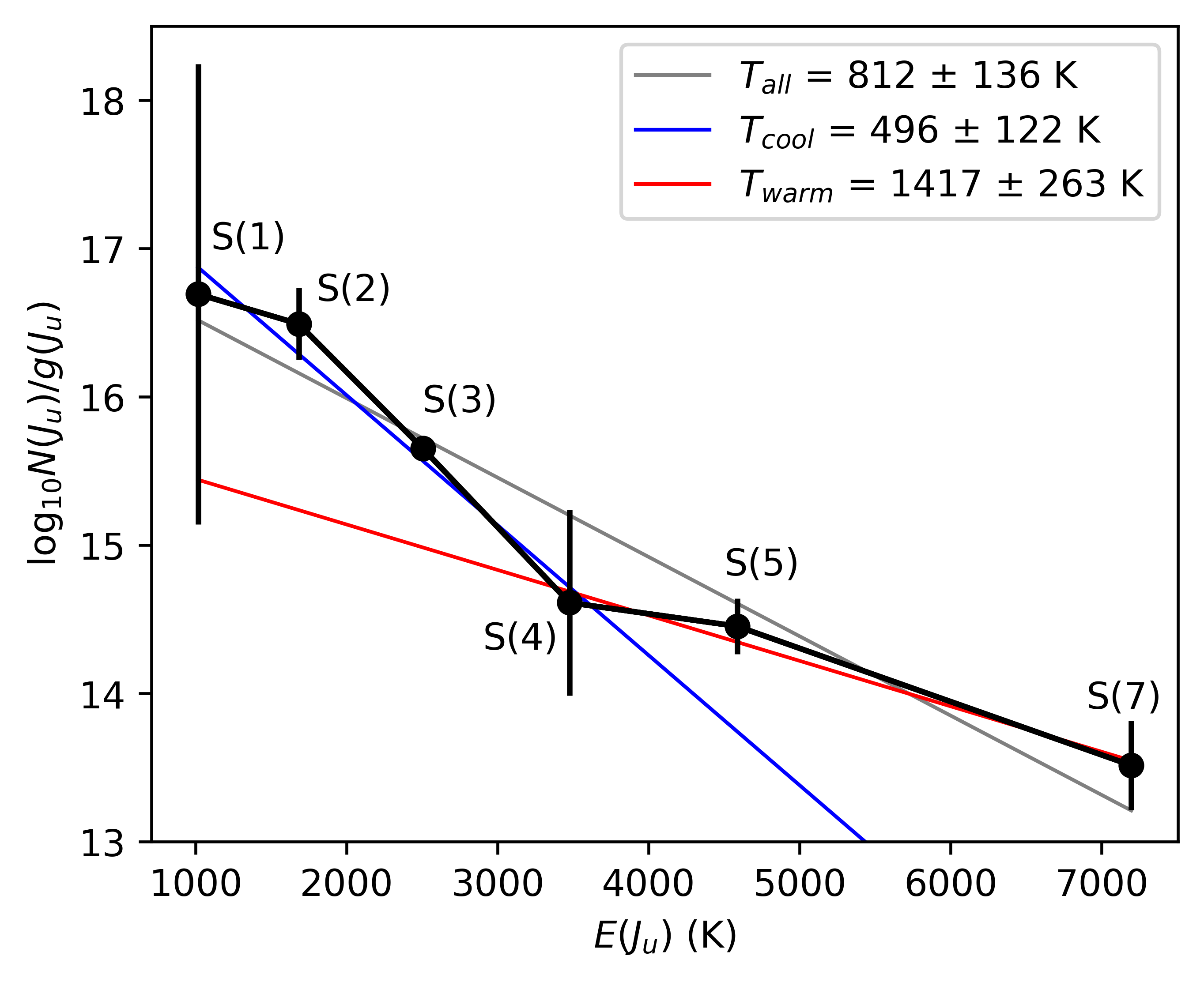}
\caption{Excitation diagram for the rotational transitions of H$_2$ in region C. Values of $g(J_u)$, $E(J_u)$, and $A$ were gathered from \cite{Rou2019} in order to convert from line intensities to $N(J_u)$. We fit the data points with a linear regression to determine the temperature of the gas based on the slope. Three separate fits are shown: all of the H$_2$ lines (gray line), the cooler H$_2$ tracers (S(1), S(2), S(3), S(4); blue line), and the warmer H$_2$ tracers (S(4), S(5), S(7); red line).}
    \label{fig:temperature}
\end{figure}

The S/N is not sufficient enough to build the H$_2$ excitation diagram at every position in the outflow. Assuming that the H$_2$ excitation conditions are similar throughout the outflow, we scale the excitation diagram in Figure \ref{fig:temperature} to the observed total flux in the outflow in H$_2$ $0-0$ S(3), the best resolved H$_2$ $0-0$ transition. Then we use the fit $N(0,0)$ scaled up by the partition function $Z(T)$ \citep{Her1996} to get $N_{H_2}$ and multiply by the total area subtended by outflowing H$_2$ $0-0$ S(3) to get a mass of the outflowing warm molecular gas of M$_{H_2}$ = $2.7_{-0.4}^{+0.4}$ $\times$ 10$^5$ $M_{\odot}$.

We then use Equations \ref{eqn:massoutflow} $-$ \ref{eqn:energyoutflow} and follow the same methods as the \neiii\ derivation to get the mass, momentum, and energy outflow rates for warm H$_2$ and present them in Table \ref{tab:outlflow}. It should be noted that we are unable to probe molecular gas cooler than $\sim$ 500 K as the H$_2$ $0-0$ S(0) 28.22 line lies outside of the MIRI wavelength range. This makes our estimates of M$_{H_2}$ a lower limit as most of the molecular gas appears to be in cooler phases.

Our derived temperatures of the warm molecular gas in PDS 456 are consistent with those derived from other sources showing emission of rotational excited molecular hydrogen, including starburst galaxies \citep{Rig2002, Bei2015}, Seyfert Galaxies \citep{Rig2002, Alv2023}, and  UltraLuminous InfraRed Galaxies \citep[ULIRGs,][]{Hig2006}. Both AGN and starburst can thermally excite H$_2$ through UV and X-ray radiation and/or shocks caused by the outflow \citep{Bei2015}. It is expected that H$_2$ emission should be stronger in AGN-dominated sources due to the increased X-ray emission \citep{Tin1997, Rig2002}. This is typically measured when comparing the H$_2$/PAH ratios in the extended emission but with our current data, we see no clear evidence of extended PAH emission so we cannot make this measurement. The ratio of warm to cool molecular gas is $M_{H_2}^{\mathrm{warm}}$/$M_{H_2}^{\mathrm{cool}}\sim 10^{-3}$, where $M_{H_2}^{\mathrm{cool}}$ is based on the CO(3-2) measurements of \citet{Bis2019}. This ratio is fairly low given this source is AGN-dominated but there is a high scatter in the relationship \citep{Rig2002} and we lack emission from the S(0) line, which is sensitive to the cooler warm-H$_2$ gas and would raise the value of $M_{H_2}^{\mathrm{warm}}$. 

\subsubsection{Putting it all Together}
\label{subsubsec:outflow_overview}

\begin{figure}
\centering
\includegraphics[width=0.45\textwidth]{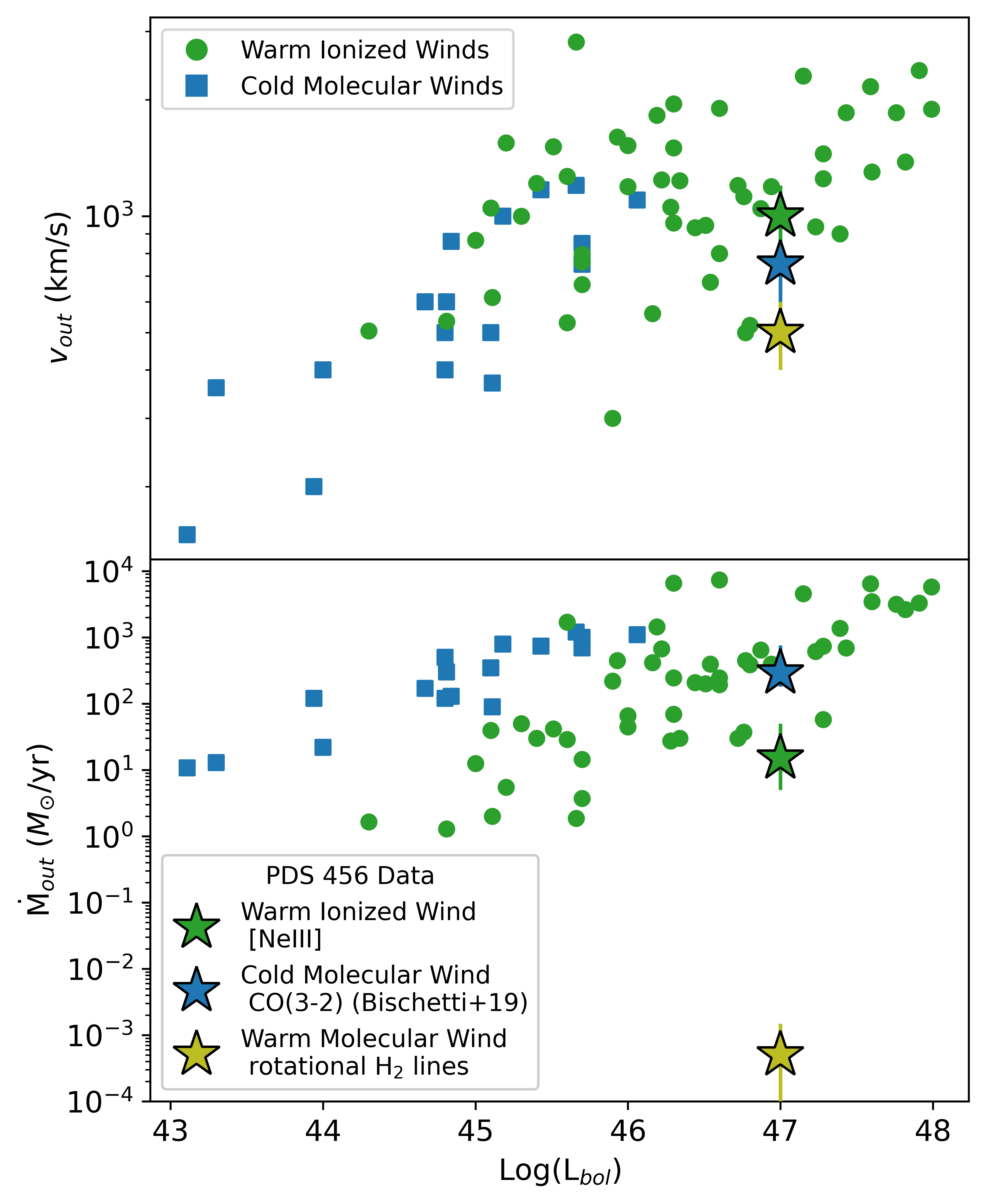}
\caption{Comparison of the outflow kinematics of PDS 456 to those of other AGN-driven outflows. Data for cold molecular winds (blue) and warm ionized winds (green) are taken from \cite{Fio2017}. For PDS 456, the cold molecular wind measurements (blue star) are based on the CO(3-2) measurements of \citet{Bis2019}, the warm ionized wind measurements (green star) are calculated from the \neiii\ luminosities, and the warm molecular wind measurements (yellow star) are calculated from the MIR rotational H$_2$ lines. See text for more details.}
    \label{fig:cosmicnoon}
\end{figure}

In Figure \ref{fig:cosmicnoon}, we compare the measured outflow parameters, $v_{out}$ and $\dot M_{\mathrm{out}}$, for each outflow phase of PDS 456, with data from other AGN-driven outflows \citep{Fio2017}. In general, we see that the outflow velocities and mass outflow rates of PDS 456 lie on the lower end of the scatter of the data tabulated in \cite{Fio2017}, but are not wholly inconsistent with the relations. For PDS 456, the cold molecular gas is the dominant phase of the outflowing gas, contributing $\sim95\%$ to the total mass outflow rate. The warm ionized gas contributes $\sim5\%$ to the total mass outflow rate and the warm molecular gas contributes $\ll 1\%$. 

Putting all of these individual phases together, we compare the calculated energetics of the multiphase outflow to the energetics of the radiation from the central quasar, $\dot P_{rad} = L_{bol}/c$ and $\dot E_{rad} = L_{bol} = 10^{47}$ \ergs. We get a momentum boost of $(\dot P_{H_2, warm} + \dot P_{H_2, cool} + \dot P_{ion})/\dot P_{rad}\sim0.39$, and a ratio of the kinetic energy outflow rate to the AGN luminosity of $\dot E_{out}/\dot E_{rad}$ $\sim$ 5 $\times$ $10^{-4}$. These values show that the quasar supplies enough momentum and energy to power the outflow and are consistent with model predictions for a relatively efficient momentum-conserving radiatively driven outflow \citep[$\dot P_{out}/\dot P_{rad}\sim$1;][]{Fab1999, Kin2003, Cos2018}. The momentum boost is consistent, but slightly larger than the value reported by \cite{Tra2024}, who used \oiii\ instead of \neiii\ to trace the warm ionized gas and lacked the contribution from the warm molecular gas (which does not significantly impact the results).  

For completeness, we also discuss two other outflow mechanisms: energy-driven outflow and starburst-driven outflow. An energy-driven outflow typically results in a high momentum boost, perhaps as high as $\sim10-20$ \citep{Zub2012, Fau2012}, which is inconsistent with our results. \cite{Tra2024} reconcile this with an intermittent AGN phase scenario where the long-term average $L_{bol}$ of PDS 456 is lower than its current state by $\sim 20 \%$. This scenario cannot be ruled out as, at the current outflow velocities ($200$ \kms), it would take $30$ Myr for gas to reach the edge of the outflow at 6 kpc. This is longer than typical AGN variability timescales of $\sim 0.1$ Myr \citep{Sch2015} and comparable to the estimated lifetimes of luminous quasars \citep{Mar2004}. A starburst-driven outflow was excluded by \cite{Bis2019} from a comparison of the molecular outflow to the calculated SFR. Due to our agreement with their SFR and our calculation of a high AGN contribution of 93\%, we also disfavour a purely starburst-driven outflow.

\section{Conclusion}
\label{sec:conclusion}

We uniformly analyzed the MIRI, NIRSpec, and MUSE IFS data of the $z$ = 0.185 luminous quasar PDS 456. The new \jwst\ data provide the first high-resolution infrared view of the outflow and host galaxy around this extremely luminous quasar ($L_{bol} \sim 10^{47}$ \ergs). We focus our analysis on the MIRI data but use the high-S/N MUSE and NIRSpec data to help frame our analysis of the more noisy MIRI data. For all three data sets, we use \qtdfit\ to separate the quasar light from the host galaxy to derive maps of emission line fluxes and kinematics. We also analyze the nuclear spectrum extracted from the MIRI data cube. The main results of our analysis include: 

\begin{enumerate}
    \item Both the NIRSpec and MUSE data cubes show significant evidence of an eastern outflow extending out to $\sim 15$ kpc from the quasar in the MUSE data, and at least $\sim 8$ kpc in the NIRSpec data (limited by the FOV of NIRSpec). This outflow is seen in both the warm-ionized (\oiii, \paa) and warm-molecular (H$_2$ 1-0 S(3)) gas tracers. Both gas phases have similar kinematics with maximum outflow velocities $v_{90} \sim -400$ \kms. Our analysis adds evidence of a multiphase outflowing shell to the east of PDS 456, nearly perpendicular to the rotating cold-molecular disk seen in the published ALMA data.  
    
    \item Our analysis of the MIRI data cube detects this same outflow in several mid-infrared emission lines spanning a range of ionization potentials and temperatures. These emission lines show morphologies and kinematics that are similar to those traced by the optical and near-infrared lines in the MUSE and NIRSpec data cubes, respectively. The warm ionized gas component of the outflow is highly ionized, up to \nev\ and \nevi, but is not clearly detected in lower ionization lines such as \neii\ and \arii. There is also a significant detection of warm-molecular H$_2$ $0-0$ line emission, roughly co-spatial and sharing the same kinematics as the warm ionized gas. It is a spectacular confirmation of the multi-phase nature of the outflow in this quasar. 

    \item A more in-depth analysis of these data reveals that this multi-phase outflow involves a warm ionized gas mass of $M_{\mathrm{ion}}$ = 2.44$_{-0.6}^{+0.6}\times 10^7$ $M_{\odot}$ and a warm molecular gas mass of $M_{H_2}^{\mathrm{warm}}$ = 2.7$_{-0.4}^{+0.4}\times 10^5$ $M_{\odot}$. These represent respectively only $\sim$ 10\% and $\sim$ 0.1\% of the mass of outflowing cold molecular gas detected in the ALMA data. The total outflow momentum rate from all three gas phases imply a modest momentum boost of $\dot P_{out}/\dot P_{rad}\sim0.39$, which is consistent with a radiatively driven outflow.  
    
    \item Our detailed analysis of the MIRI nuclear spectrum of PDS 456 shows that the central quasar contributes 93\% of the bolometric luminosity. In the nuclear region, we detect PAH emission indicative of a star formation rate of 39 M$_{\odot}$ yr$^{-1}$ and dust grain size distribution on average smaller than that in Seyfert galaxies, although the uncertainties on the PAH ratios are significant. In addition to driving the wind, we find that this luminous quasar interacts with its host galaxy in at least two other ways. The AGN radiation excites silicates in dust clouds 300 pc from the source and dominates the ionization structure of the host galaxy and outflow.

\end{enumerate}

Our analysis of the new \jwst\ IFS data on PDS 456 expands our understanding of the multi-phase outflow in this extreme quasar. These data clearly reveal the impact of such a powerful central energy source on its host galaxy in all gas and dust phases. This source is a fantastic nearby analog to quasars at the peak of supermassive black hole accretion around $z\sim 2$. However, this bright source is pushing the limits of \jwst\ and MIRI/MRS. The undersampled PSF limits the reliability of the analysis of the MIRI data. Future improvements to the MIRI MRS data reduction pipeline should eventually make it possible to better constrain the full extent of quasar feedback in this system. 


\begin{acknowledgments}

J.S. and S.V.\ acknowledge partial financial support by NASA for this research through STScI grants No.\  JWST-ERS-01335, JWST-GO-01865, JWST-GO-02547, JWST GO-03869, and JWST GO-05627.  W.L., D.S.N.R., A.V., and N.L.Z. were also partially supported by grant JWST-ERS-01335. 

\end{acknowledgments}

%

\vspace{5mm}


\software{astropy \citep{Ast2013, Ast2018},  \qtdfit\ \citep{Rup2023}}




\bibliography{pdsbib}{}
\bibliographystyle{aasjournal}



\end{document}
